\documentclass[onecolumn, ceqn, showpacs, showkeys, prd, onecolumn, byrevtex, nofootinbib, amsmath, amsfonts, amssymb]{revtex4-1}

\usepackage{amssymb,amsmath,mathtools,xcolor,graphicx,xspace,colortbl,ragged2e,rotating} % 
\usepackage{textcomp}
\usepackage{amsfonts}  %% 010
\usepackage{amsmath}  %% 010
\usepackage{amssymb}  %% 100
\usepackage{boxedminipage}  %% 100
\usepackage{graphics}  %% 100
\usepackage{graphicx}  %% 100
\usepackage{ragged2e}  %% 100
\usepackage{tabulary}  %% 100
\usepackage{varioref}  %% 100
\usepackage{wrapfig}  %% 100
\usepackage{xcolor}  %% 200
\usepackage{url}
\graphicspath{{FRACTIONAL3_1_graphics/}{FRACTIONAL3_1_tcache/}{FRACTIONAL3_1_gcache/}}
\DeclareGraphicsExtensions{.pdf,.eps,.ps,.png,.jpg,.jpeg}

\begin{document}\thinspace

\title{NEWTONIAN FRACTIONAL-DIMENSION GRAVITY AND DISK GALAXIES
}
\author{Gabriele U. Varieschi
}
\email[E-mail me at: ]{gvarieschi@lmu.edu
}
\homepage[Visit: ]{http://gvarieschi.lmu.build
}
\affiliation{Loyola Marymount University, Los Angeles, CA 90045, USA
}

\date{
%TCIMACRO{\TeXButton{today}{\today}}%
%BeginExpansion
\today
%EndExpansion
}
\begin{abstract}
This paper continues previous work on a novel alternative model of gravity, based on the theory of fractional-dimension spaces applied to Newton's law of gravitation. In particular, our Newtonian Fractional-Dimension Gravity is now applied to axially-symmetric structures, such as thin/thick disk galaxies described by exponential, Kuzmin, or other similar mass distributions.

As in the case of spherically-symmetric structures, which was studied in previous work on the subject, we examine a possible connection between our model and Modified Newtonian Dynamics, a leading alternative
gravity model, which accounts for the observed properties of galaxies and other astrophysical
structures without requiring the dark matter hypothesis.

By relating the MOND acceleration constant
$a_{0} \simeq 1.2 \times 10^{ -10}\mbox{m}\thinspace \mbox{s}^{ -2}$
to a natural scale length
$l_{0}$
of our model, namely
$a_{0} \approx GM/l_{0}^{2}$ for a galaxy of mass
$M$, and by using the empirical Radial Acceleration
Relation, we are able to explain the connection between the observed radial acceleration
$g_{obs}$ and the baryonic radial acceleration
$g_{bar}$ in terms of a variable local dimension $D$. As an example of this methodology, we provide a detailed rotation curve fitting for the case of the field dwarf spiral galaxy NGC 6503.
\end{abstract}
\maketitle

\section{\label{sect:intro}
Introduction
}
 This paper continues the discussion introduced in a previous publication (\cite{Varieschi:2020ioh}, paper I in the following) of a possible \textit{Newtonian Fractional-Dimension Gravity} (NFDG), an extension of the standard laws of Newtonian gravity to lower dimensional spaces, including those with fractional
(i.e., non-integer) dimension (see also Ref. \cite{Varieschi:2020} for a general introduction to NFDG). 

This model is loosely based on the methods of fractional mechanics and fractional electromagnetism (see \cite{bookTarasov,bookZubair,Varieschi:2018}, and references therein) and on the general framework of fractional calculus
(FC) \cite{MR0361633,MR1219954,MR1658022,Herrmann:2011zza,MR1890104}. Fractional calculus is also usually related to
fractal geometries, which might play a role at galactic or cosmological scales in the universe \cite{Baryshev:2002tn,Nottale:2011zz,Calcagni:2016azd}. The idea of using fractional calculus to modify gravity is not new, as it appears in many papers in the literature  \cite{Muslih2007,Rousan2002,Munkhammar:2010gq,Calcagni:2009kc,Calcagni:2011kn,Calcagni:2011sz,Calcagni:2013yqa,Calcagni:2016azd,Calcagni:2016xtk,Calcagni:2018dhp,Svozil:2017ybx,Giusti:2020rul,Giusti:2020kcv}, just to mention only a few. 

However, as already discussed in detail in our paper I, we note that NFDG is not a fractional theory in the sense used by most of the other models in the above references. The ``fractional'' equations for the gravitational potential and field that will be introduced in Sect. \ref{sect:REVIEW} are based on operators constructed with ordinary derivatives, not fractional ones. Thus, our model has a fractal structure, due to the non-integer dimension of the metric space, and would be better described as ``Newtonian gravity in fractional dimensional spaces.'' We prefer to call it NFDG for simplicity's sake, as it was done in paper I. 

In paper I, we developed the bases of this generalized NFDG, with focus on spherically-symmetric structures. In this work, we extend our analysis to axially-symmetric structures, and in particular to the case of thin/thick disk galaxies. A possible connection of NFDG with Modified Newtonian Dynamics
(MOND) (\cite{Milgrom:1983ca,Milgrom:1983pn,Milgrom:1983zz} and \cite{2002ARA&A..40..263S,Famaey:2011kh,2014SchpJ...931410M,Merritt:2020pwe} for MOND reviews) was also shown in our paper I. The strong empirical correlation between the radial gravitational acceleration traced by galactic
rotation curves and that predicted by using the observed distribution of baryons (Radial Acceleration Relation - RAR \cite{McGaugh:2016leg,Lelli:2017vgz,Chae:2020omu}) was explained in terms of a variable dimension $D$ in galactic structures \cite{Varieschi:2020ioh}, thus connecting MOND with
NFDG.

In Sect. \ref{sect:REVIEW}, we review the basic ideas of NFDG from paper I, and the connections with MOND and the RAR. In Sect. \ref{sect:AXIAL}, we
extend NFDG to axially-symmetric cases, by considering models for thin/thick disk galactic structures. In Sect. \ref{sect::galactic}, we apply these methods to simple mass distributions, such as exponential and Kuzmin disks, and also briefly revisit the case of spherical structures. In Sect. \ref{sect:6503}, we examine the particular case of the field dwarf spiral galaxy NGC 6503, as an example of our methods. Finally, in
Sect. \ref{sect::conclusion} conclusions are drawn
and possible future work on the subject is outlined.

\section{\label{sect:REVIEW} Review of NFDG and its connections with MOND
}
Newtonian Fractional-Dimension Gravity \cite{Varieschi:2020ioh} was introduced heuristically by considering extensions of Gauss's law for gravitation to a lower-dimensional space-time $D +1$, where $D$ can be a non-integer space dimension. In particular, this model was based on the original dimensional regularization techniques used in
quantum field theory \cite{Bollini:1972ui,tHooft:1972tcz,Wilson:1972cf}. Similar regularization
methods \cite{1987JPhA...20.3861S,bookTarasov,bookZubair} were used to generalize the integral of spherically-symmetric functions over a $D$-dimensional metric space and the axiomatic bases for spaces with non-integer dimension
were introduced by Stillinger \cite{doi:10.1063/1.523395} and Wilson
\cite{Wilson:1972cf}, and later refined by Palmer and
Stavrinou \cite{Palmer_2004}. 

The gravitational field of a point-like mass was determined as \cite{Varieschi:2020ioh}:

\begin{equation}\left \vert \mathbf{g}\right \vert  =2\pi ^{1 -D/2}\Gamma (D/2)\frac{G\widetilde{m}_{(D)}}{l_{0}^{2}}\frac{1}{(r/l_{0})^{D -1}} , \label{eq2.1}
\end{equation}where the scale length $l_{0}$ is needed to ensure dimensional correctness of the expression for $D \neq 3$ and $\widetilde{m}_{\left (D\right )}$ represents a D-dimensional point mass (see paper I for details).

Since $D$ can assume non-integer values, it is convenient to use dimensionless coordinates in all formulas, starting with the radial distance $w_{r} \equiv r/l_{0}$ or, in general,  introducing dimensionless coordinates
$\mathbf{w} \equiv \mathbf{x}/l_{0}$
for the field point and
$\mathbf{w}^{ \prime } \equiv \mathbf{x}^{ \prime }/l_{0}$
for the source point. We also introduce a rescaled mass ``density''
$\widetilde{\rho }\left (\mathbf{w}^{ \prime }\right ) =\rho \left (\mathbf{w}\mathbf{}^{ \prime }l_{0}\right )l_{0}^{3} =\rho \left (\mathbf{x}^{ \prime }\right )l_{0}^{3}$, where
$\rho (\mathbf{x}^{ \prime })$
is the standard mass density in
$\mbox{kg}\thinspace \mbox{m}^{ -3}$, so that $d\widetilde{m}_{\left (D\right )} =\widetilde{\rho }\left (\mathbf{w}^{ \prime }\right )d^{D}\mathbf{w}^{ \prime }$ represents the infinitesimal source mass and Eq. (\ref{eq2.1}) can be generalized to a mass distribution over the D-dimensional source volume
$V_{D}$ as follows \cite{Varieschi:2020ioh}: \begin{equation}\mathbf{g}(\mathbf{w}) = -\frac{2\pi ^{1 -D/2}\Gamma (D/2)G}{l_{0}^{2}}{\displaystyle\int _{V_{D}}}\widetilde{\rho }(\mathbf{w}^{ \prime })\frac{\mathbf{w} -\mathbf{w}^{ \prime }}{\left \vert \mathbf{w} -\mathbf{w}^{ \prime }\right .\vert ^{D}}d^{D}\mathbf{w}^{ \prime } . \label{eq2.2}
\end{equation}

As in standard Newtonian gravity, a gravitational potential
$\phi \left (\mathbf{w}\right )$
was introduced as \cite{Varieschi:2020ioh}:
\begin{gather}\phi (\mathbf{w}) = -\frac{2\pi ^{1 -D/2}\Gamma (D/2)G}{\left (D -2\right )l_{0}^{2}}{\displaystyle\int _{V_{D}}}\frac{\widetilde{\rho }(\mathbf{w}^{ \prime })}{\left \vert \mathbf{w} -\mathbf{w}^{ \prime }\right .\vert ^{D -2}}d^{D}\mathbf{w}^{ \prime };\ D \neq 2 \label{eq2.3} \\
\phi \left (\mathbf{w}\right ) =\frac{2G}{l_{0}^{2}}{\displaystyle\int _{V_{2}}}\widetilde{\rho }\left (\mathbf{w}^{ \prime }\right )\ln \left \vert \mathbf{w} -\mathbf{w}^{ \prime }\right .\vert d^{2}\mathbf{w}^{ \prime };\ D =2 \nonumber \end{gather}
with
$\phi (\mathbf{w})$
and
$\mathbf{g}(\mathbf{w})$
connected by
$\mathbf{g}(\mathbf{w}) = - \nabla _{D}\phi (\mathbf{w})$, where the D-dimensional gradient
$ \nabla _{D}$
is considered equivalent to the standard one. It is easy to check that all the expressions in Eqs. (\ref{eq2.1})-(\ref{eq2.3}) above, correctly reduce to the standard Newtonian ones for $D =3$.\protect\footnote{
Since
$ \nabla _{D}$ is defined in terms
of dimensionless coordinates, the physical dimensions for the gravitational potential
$\phi $ in Eq. (\ref{eq2.3}) are the same as
those for the gravitational field
$\mathbf{g}$, i.e., both quantities will be measured in $\mbox{m}\thinspace \mbox{s}^{ -2}$. Therefore, the Newtonian potential is obtained as $\phi _{Newt} =l_{0}\phi _{D =3}\left (\mathbf{w}\right )$, from the NFDG potential in the first line of Eq. (\ref{eq2.3}) for fixed $D =3$.}

The gravitational potential in Eq. (\ref{eq2.3}) was also suggested by the solutions to the D-dimensional Laplace equation in spherical coordinates and the related multipole expansion \cite{Varieschi:2020ioh}. All these expressions were derived for a fixed value of the (fractional) dimension $D$, but we argued that they are approximately valid also in the case of a variable dimension $D\left (\mathbf{w}\right )$, assuming a slow change of the dimension $D$ with the field point coordinates.

The scale length $l_{0}$ is needed to connect our expressions in Eq. (\ref{eq2.2}), or Eq. (\ref{eq2.3}), with the physical reality. In paper \ I, we argued that this scale length might be related to the MOND\  acceleration constant
$a_{0}$, whose currently estimated value is also denoted by $g_{\dag }$ \cite{McGaugh:2016leg,Lelli:2017vgz}:

\begin{equation}a_{0} \equiv g_{\dag } =1.20 \pm 0.02\ \text{(random)} \pm 0.24\ \text{(syst)} \times 10^{ -10}\ \mbox{}\ \mbox{m}\thinspace \mbox{s}^{ -2} , \label{eq2.4}
\end{equation}
and which represents the acceleration scale below which MOND corrections are
applied.

MOND \cite{Milgrom:1983ca,Milgrom:1983pn,Milgrom:1983zz} modifies Newtonian dynamics in two possible ways
\cite{Bekenstein:1984tv}:

\begin{gather}m\mu (a/a_{0})\mathbf{a} =\mathbf{F} \label{eq2.5} \\
\mu (g/a_{0})\mathbf{g} =\mathbf{g}_{N} , \nonumber \end{gather}
where the former indicates modified inertia (MI), since the mass $m$ is replaced by $m\mu \left (a/a_{0}\right )$, while the latter indicates modified gravity (MG), since the observed gravitational field $\mathbf{g}$ can differ from the Newtonian one, $\mathbf{g}_{N}$. The two formulations are practically equivalent, but conceptually
different:\ the former modifies Newton's laws of motion, while
the latter modifies Newton's law of universal gravitation.

There is now limited evidence \cite{Petersen:2020vks,Milgrom:2012rk} that MG might be favored over MI, according to preliminary studies of galactic rotation curves which might be able to differentiate between the two models. In view of Eqs. (\ref{eq2.2})-(\ref{eq2.3}) above, our NFDG should also be considered a modification of the law of gravity, since we assume that a test object, subject to the fractional gravitational field described by Eq. (\ref{eq2.2}), will still move in a
(classical)
$3 +1$
space-time, thus obeying standard laws of dynamics. 

However, in both MI and MG interpretations the modifications of the Newtonian laws follow
from the \textit{interpolation function }$\mu (x) \equiv \mu (a/a_{0})\text{}$
or
$\mu \text{}(x) \equiv \mu (g/a_{0})$, respectively. MOND assumes that:

\begin{equation}\mu \left (x\right ) \approx \left \{\begin{array}{c}1\text{  for }x \gg 1\text{  (Newtonian regime)}\text{
} \\
x\text{  for }x \ll 1\text{  (Deep-MOND limit)}\text{
}\end{array}\right \} \label{eq2.6}
\end{equation}
and it has become customary \cite{McGaugh:2008nc} to
substitute the interpolation function $\mu \left (x\right )$ with its inverse function $\nu \left (y\right )$, i.e.: $\mathbf{g} =\mu ^{ -1}(x)\mathbf{g}_{N} \equiv \nu (y)\mathbf{g}_{N}\ \text{with}\ y =g_{N}/a_{0}$.

As it was done in paper I, we will consider two main families of
$\nu (y)$
functions \cite{McGaugh:2008nc}:

\begin{gather}\nu _{n}(y) =\left (\frac{1}{2} +\frac{1}{2}\sqrt{1 +4y^{ -n}}\right )^{1/n} \label{eq2.7} \\
\widehat{\nu }_{n}(y) =\left [1 -\exp \left ( -y^{n/2}\right )\right ]^{ -1/n} , \nonumber \end{gather}
where the particular choice
$\widehat{\nu }_{1}(y) =\left [1 -\exp \left ( -y^{1/2}\right )\right ]^{ -1}$
has recently become the favorite interpolation function \cite{McGaugh:2016leg,Lelli:2017vgz,Chae:2020omu}. This function is equivalent to the so-called Radial Acceleration Relation - RAR:

\begin{equation}g_{obs} =\frac{g_{bar}}{1 -e^{ -\sqrt{g_{bar}/g_{\dag }}}} , \label{eq2.8}
\end{equation}
where
$g_{\dag }$
is an empirical parameter corresponding to the MOND
acceleration scale
$a_{0}$, as already reported in Eq. (\ref{eq2.4})
above, and we also identify$\ y =g_{N}/a_{0} \equiv g_{bar}/g_{\dag }$ ,
$\widehat{\nu }_{1} =g/g_{N} \equiv g_{obs}/g_{bar}$.

Currently, the RAR represents the best empirical fit \cite{McGaugh:2016leg,Lelli:2017vgz,Chae:2020omu} relating the
radial acceleration
$g_{obs}$
traced by rotation curves with the radial acceleration
$g_{bar}$
predicted by the observed distribution of baryonic matter in galaxies and was obtained by using data points from a sample of 175 galaxies in the Spitzer Photometry and Accurate Rotation Curves
(SPARC) database \cite{Lelli:2016zqa}. 

This empirical relation was confirmed in more recent work  \cite{Lelli:2017vgz}, by adding
early-type-galaxies (elliptical and lenticular) and dwarf spheroidal galaxies to the SPARC database, and also \cite{Li:2018tdo} by checking individually the 175 galaxies in the original SPARC catalogue against the RAR, allowing for galaxy-to-galaxy variations of the acceleration scale
$g_{\dag }$. The result of this analysis still favors a single value of
$g_{\dag }$, consistent with the action of a single effective force law. In addition, a detection of the external field effect, which is typical of Milgromian dynamics (MOND), has been recently reported \cite{Chae:2020omu}.

Heuristically, in paper I we proposed a possible connection between the scale length $l_{0}$ and the MOND\ acceleration $a_{0}$ as:
\begin{equation}l_{0} \approx C\sqrt{\frac{GM}{a_{0}} ,} \label{eq2.9}
\end{equation}where $C >0$ is a constant and $M$ is the total mass (or a reference mass) of the system being studied.

Assuming, for simplicity's sake, $C =1$; therefore,

\begin{equation}a_{0} \approx \frac{GM}{l_{0}^{2}} , \label{eq2.10}
\end{equation}we were able to show in paper I\ that the main consequences of the MOND\ theory could be recovered from Eq. (\ref{eq2.1}) by considering the Deep-MOND Limit (DML) equivalent to reducing the space dimension to $D \approx 2$. In particular, the asymptotic or
flat rotation velocity
$V_{f} \approx \sqrt[{4}]{GMa_{0}}$ shown by galactic rotation curves, the ``baryonic'' Tully-Fisher relation-BTFR:
$M_{bar} \sim V_{f}^{4}$, and other fundamental MOND predictions were recovered with our NFDG for the case $D \approx 2$  \cite{Varieschi:2020ioh}.\protect\footnote{
Actually, MOND predictions were recovered for any positive value for the constant $C$ in equation (\ref{eq2.9}), showing that $M$ can be considered as an arbitrary reference mass of the galactic structure being studied in our model.
}

It should be noted that our NFDG, as well as other similar models recently introduced (\cite{Giusti:2020rul,Giusti:2020kcv}, see also related discussion in paper I), is not a fractional or a fractal version of MOND, but rather a MOND-like model reproducing the asymptotic behavior of MOND. In fact, MOND is a fully non-linear theory, while NFDG is inherently linear, in view of its fundamental equations presented in this section. However, in this work as well as in paper I, we also attempt to describe the transition between the two asymptotic regimes of MOND by assuming a continuous (slow) change in the variable dimension of the associated metric space.

To conclude this review of the main findings of our paper I, for spherically-symmetric mass distributions $\widetilde{\rho }\left (w_{r}^{ \prime }\right )$ we were able to prove that the gravitational field
$\mathbf{g}(w_{r})$, in a fractal space of dimension
$D(w_{r}^{})$, depending on the radial distance $w_{r} =r/l_{0}$ from the center of the coordinate system, can be computed
as:

\begin{equation}\mathbf{g}_{obs}(w_{r}) = -\frac{4\pi G}{l_{0}^{2}w_{r}^{D\left (w_{r}\right ) -1}}{\displaystyle\int _{0}^{w_{r}}}\tilde{\rho }\left (w_{r}^{ \prime }\right )w_{r}^{ \prime ^{D\left (w_{r}\right ) -1}}dw_{r}^{ \prime }\overset{}{\widehat{\mathbf{w}}_{r} ,} \label{eq2.11}
\end{equation}
for
$1 \leq D \leq 3$. In the previous equation, we also denoted the gravitational field as the
``observed'' one,
$\mathbf{g}_{obs}$, as opposed to the ``baryonic''
$\mathbf{g}_{bar}$:

\begin{equation}\mathbf{g}_{bar}(w_{r}) = -\frac{4\pi G}{l_{0}^{2}w_{r}^{2}}{\displaystyle\int _{0}^{w_{r}}}\tilde{\rho }\left (w_{r}^{ \prime }\right )w_{r}^{ \prime ^{2}}dw_{r}^{ \prime }\overset{}{\widehat{\mathbf{w}}_{r} ,} \label{eq2.12}
\end{equation}
for fixed dimension
$D =3$. Therefore, we identified the observed and baryonic accelerations
$g_{obs}$
and
$g_{bar}$ \cite{McGaugh:2016leg} with those obtained in NFDG for
variable dimension
$D\left (w_{r}\right )$
and for fixed dimension
$D =3$, respectively.

 With these NFDG assumptions, and for spherically symmetric structures, the ratio
$(g_{obs}/g_{bar})_{NFDG}$ was simply obtained from Eqs. (\ref{eq2.11}) and (\ref{eq2.12}):

\begin{equation}\genfrac{(}{)}{}{}{g_{obs}}{g_{bar}}_{NFDG}(w_{r}) =w_{r}^{3 -D\left (w_{r}\right )}\frac{{\displaystyle\int _{0}^{w_{r}}}\widetilde{\rho }\left (w_{r}^{ \prime }\right )w_{r}^{ \prime ^{D\left (w_{r}\right ) -1}}dw_{r}^{ \prime }}{{\displaystyle\int _{0}^{w_{r}}}\widetilde{\rho }\left (w_{r}^{ \prime })\right .w_{r}^{ \prime ^{2}}dw_{r}^{ \prime }} , \label{eq2.13}
\end{equation}
where the dimension function
$D(w_{r})$ was computed in paper I by comparing the expression in Eq. (\ref{eq2.13}) with the MOND-RAR
equivalent expression from Eq. (\ref{eq2.8}),
$\genfrac{(}{)}{}{}{g_{obs}}{g_{bar}}_{MOND}\left (w_{r}\right ) =\frac{1}{1 -e^{ -\sqrt{g_{bar}\left (w_{r}\right )/g_{\dag }}}}$, or with similar expressions obtained by using the other interpolation functions in Eq. (\ref{eq2.7}).

In fact, this procedure was successful for several different forms of the spherically-symmetric mass distributions analyzed in paper I. For each case, the computed dimension functions
$D(w_{r}^{})$ assumed values
$D \approx 3$
in regions where Newtonian gravity was known to hold. The dimension was decreasing continuously
toward $D \approx 2$ in regions where the DML applied, following our general discussion of NFDG outlined above. In the next sections, we will show how a similar analysis can be performed for axially-symmetric mass distributions, in particular for thin/thick disk galactic structures.

\section{\label{sect:AXIAL} NFDG and axially-symmetric mass distributions
}
Adapting NFDG to axially-symmetric mass distributions presents some mathematical challenges. Rather than computing directly the gravitational field using Eq. (\ref{eq2.2}), as it was done for the spherically-symmetric case \cite{Varieschi:2020ioh}, we prefer to calculate the gravitational potential using Eq. (\ref{eq2.3}) and then apply the fractional gradient
$ \nabla _{D}$, which is equivalent to the standard one.

Before we follow this procedure, we recall that, in the standard $D =3$ case, different methods \cite{2008gady.book.....B} are used to evaluate axisymmetric potentials and fields. One of the most popular options in alternative theories of gravity, such as in Mannheim's Conformal Gravity \cite{Mannheim:2005bfa}, is to make use of the cylindrical coordinate Green's function expansion in terms of Bessel functions of the first kind $J_{\nu }$ (\cite{Jackson:1998nia}, problem 3.16):

\begin{equation}\frac{1}{\left \vert \mathbf{r} -\mathbf{r}^{ \prime }\right \vert } =\sum \limits _{m = -\infty }^{\infty }\int _{0}^{\infty }dkJ_{m}\left (kR\right )J_{m}\left (kR^{ \prime }\right )e^{im\left (\varphi  -\varphi ^{ \prime }\right ) -k\left \vert z -z^{ \prime }\right \vert } , \label{eq3.1}
\end{equation}and follow an approach originally developed by Casertano \cite{1983MNRAS.203..735C}. 

In the case of a thin-disk mass distribution with axial symmetry (using standard cylindrical coordinates $R$, $\varphi $, $z$) and for exponential disks of total mass $M$ and scale length $R_{d}$, i.e.,

\begin{gather}\rho \left (R^{ \prime } ,z^{ \prime }\right ) =\Sigma \left (R^{ \prime }\right )\delta \left (z^{ \prime }\right ) \label{eq3.2} \\
\Sigma \left (R^{ \prime }\right ) =\Sigma _{0}e^{ -R^{ \prime }/R_{d}} =\frac{M}{2\pi R_{d}^{2}}e^{ -R^{ \prime }/R_{d}} , \nonumber \end{gather}
the gravitational potential $\phi $ and the field $\mathbf{g}$ can be computed analytically in the $z =0$ plane as  \cite{Mannheim:2005bfa}:

\begin{gather}\phi \left (R\right ) = -\frac{GMR}{2R_{d}^{2}}\left [I_{0}\left (\frac{R}{2R_{d}}\right )K_{1}\genfrac{(}{)}{}{}{R}{2R_{d}} -I_{1}\genfrac{(}{)}{}{}{R}{2R_{d}}K_{0}\genfrac{(}{)}{}{}{R}{2R_{d}}\right ] \label{eq3.3} \\
\mathbf{g}\left (R\right ) = -\frac{GMR}{2R_{d}^{3}}\left [I_{0}\left (\frac{R}{2R_{d}}\right )K_{0}\genfrac{(}{)}{}{}{R}{2R_{d}} -I_{1}\genfrac{(}{)}{}{}{R}{2R_{d}}K_{1}\genfrac{(}{)}{}{}{R}{2R_{d}}\right ]\widehat{\mathbf{r}} \nonumber \end{gather}
where $I_{\nu }$ and $K_{\nu }$ are modified Bessel functions.

In particular, the second line of Eq. (\ref{eq3.3}) can be rewritten using our dimensionless cylindrical radial coordinate $w_{R} =R/l_{0}$ and the rescaled disk length $W_{d} =R_{d}/l_{0}$ as:

\begin{equation}\mathbf{g}_{bar}\left (w_{R}\right ) = -\frac{GMw_{R}}{2l_{0}^{2}W_{d}^{3}}\left [I_{0}\genfrac{(}{)}{}{}{w_{R}}{2W_{d}}K_{0}\genfrac{(}{)}{}{}{w_{R}}{2W_{d}} -I_{1}\genfrac{(}{)}{}{}{w_{R}}{2W_{d}}K_{1}\left (\frac{w_{R}}{2W_{d}}\right )\right ]\widehat{\mathbf{w}}_{R} . \label{eq3.4}
\end{equation}This equation represents the equivalent of the previous Eq. (\ref{eq2.12}) for spherical symmetry, in the case of the baryonic gravitational field (in the $z =0$ plane) of an exponential thin disk and will be used later as the equation for the standard $D =3$ case.

The Green's function expansion in Eq. (\ref{eq3.1}) can also be used in the case of linear and quadratic potentials, typical of fourth-order conformal gravity  \cite{Mannheim:2005bfa}, but cannot be used in the case of a more general NFDG potential $\phi  \sim 1/\left \vert \mathbf{w} -\mathbf{w}^{ \prime }\right \vert ^{D -2}$, such as the one in the first line of Eq. (\ref{eq2.3}). As an alternative option, we consider the simplest expansion we found in the mathematical literature \cite{2012JPhA...45n5206C} for these potentials in spherical coordinates:

\begin{equation}\frac{1}{\left \vert \mathbf{r} -\mathbf{r}^{ \prime }\right \vert ^{D -2}} =\sum \limits _{l =0}^{\infty }\frac{r_{ <}^{l}}{r_{ >}^{l +D -2}}C_{l}^{\left (\frac{D}{2} -1\right )}\left (\cos \gamma \right ) , \label{eq3.5}
\end{equation}
where $r_{ <}$ ($r_{ >}$) is the smaller (larger) of $r$ and $r^{ \prime }$, $\gamma $ is the angle between the unit vectors $\widehat{\mathbf{r}}$ and $\widehat{\mathbf{r}}^{ \prime }$, and $C_{l}^{\left (\lambda \right )}\left (x\right )$ denotes Gegenbauer polynomials (see paper I or Ref. \cite{NIST:DLMF} for general properties of these special functions).\protect\footnote{
For other possible expansions of the NFDG potential $\phi  \sim 1/\left \vert \mathbf{w} -\mathbf{w}^{ \prime }\right \vert ^{D -2}$ see also \cite{doi:10.1080/10652469.2012.761613,2013SIGMA...9..042C,2015SIGMA..11..015C}. In particular, Ref. \cite{2013SIGMA...9..042C} illustrates additional expansions of the Euler kernel $\left (z -x\right )^{ -\nu }$, in terms of Jacobi, Gegenbauer, and Chebyshev polynomials. However, these alternative expressions have proven difficult to be used in the current work. Therefore, we opted to base our analysis on the expansion in Eq. (\ref{eq3.5}).
}

The expansion in Eq. (\ref{eq3.5}) can be adapted immediately to the case of cylindrical coordinates ($R$, $\varphi $, $z$). In the case of thin disks, in the $z =z^{ \prime } =0$ plane and in the $\varphi  =0$ direction, the angle $\gamma $ is replaced by $\varphi ^{ \prime }$ and the radial spherical coordinate $r$ with the cylindrical $R$:
\begin{equation}\frac{1}{\left \vert \mathbf{r} -\mathbf{r}^{ \prime }\right \vert ^{D -2}} =\sum \limits _{l =0}^{\infty }\frac{R_{ <}^{l}}{R_{ >}^{l +D -2}}C_{l}^{\left (\frac{D}{2} -1\right )}\left (\cos \varphi ^{ \prime }\right ) , \label{eq3.6}
\end{equation}
while for thick disks, with $z^{ \prime } \neq 0$ (but still in the $z =0$ plane and $\varphi  =0$ direction) we can use the following coordinate transformations:

\begin{gather}r =R \label{eq3.7} \\
r^{ \prime } =\sqrt{R^{ \prime 2} +z^{ \prime 2}} \nonumber  \\
\cos \gamma  =\frac{R^{ \prime }\cos \varphi ^{ \prime }}{\sqrt{R^{ \prime 2} +z^{ \prime 2}}} , \nonumber \end{gather}and modify the original expansion (\ref{eq3.5}) accordingly. All these expansions can be easily written also in terms of rescaled cylindrical coordinates (defined below) and thus used as expansions of the NFDG\ kernel $1/\left \vert \mathbf{w} -\mathbf{w}^{ \prime }\right \vert ^{D -2}$.

Going back to the evaluation of the potential, in order to perform the integral in the first line of Eq. (\ref{eq2.3}) we recall the techniques outlined in paper I for multi-variable integration over a metric space $W \subset \mathbb{R}^{3}$ \cite{bookTarasov,bookZubair,Tarasov:2014fda,TARASOV2015360}. Let's assume that
$W =W_{1} \times W_{2} \times W_{3}$, where each metric set
$W_{i}$
($i =1 ,2 ,3$) has Hausdorff measure
$\mu _{i}(W_{i})$
and dimension
$\alpha _{i}$. The Hausdorff measure for the product set
$W$
can be defined as
$\mu _{H}(W) =(\mu _{1} \times \mu _{2} \times \mu _{3})(W) =\mu _{1}(W_{1})\mu _{2}(W_{2})\mu _{3}(W_{3})$
and the overall dimension is
$D =\alpha _{1} +\alpha _{2} +\alpha _{3}$. Applying Fubini's theorem we have:
\begin{gather}\int _{W}f(x_{1} ,x_{2} ,x_{3})d\mu _{H} =\int _{W_{1}}\int _{W_{2}}\int _{W_{3}}f(x_{1} ,x_{2} ,x_{3})d\mu _{1}(x_{1})d\mu _{2}(x_{2})d\mu _{3}(x_{3}) , \label{eq3.8} \\
d\mu _{i}(x_{i}) =\frac{\pi ^{\alpha _{i}/2}}{\Gamma (\alpha _{i}/2)}\left \vert x_{i}\right \vert ^{\alpha _{i} -1}dx_{i} ,\ i =1 ,2 ,3. \nonumber \end{gather}

In the previous equation, it is assumed that the spatial measure is factorized along the three topological dimensions and that the overall dimension is the sum of the individual dimensions of the sub-spaces. According to the terminology in Ref. \cite{Calcagni:2016azd}, this assumption corresponds to a multifractional geometry, where the measures in position and momentum space are all factorizable in the coordinates. Typically, multifractional theories are not Lorentz invariant because they break rotation and boost invariance \cite{Calcagni:2016azd}. Therefore, the general formula in Eq. (\ref{eq3.8}) must be adapted to each particular case being considered, in terms of the choice of coordinates and in relation to any existing symmetry of the mass distribution. 

In standard cylindrical coordinates
$(R ,\varphi  ,z)$, with $x =R\cos \varphi $, $y =R\sin \varphi $, and using the definitions
for the differentials in the second line of the last equation, we have:
$d\mu _{1}d\mu _{2}d\mu _{3} =\frac{\pi ^{\alpha _{R}/2}}{\Gamma \left (\alpha _{R}/2\right )}\frac{\pi ^{\alpha _{\varphi }/2}}{\Gamma \left (\alpha _{\varphi }/2\right )}\frac{\pi ^{\alpha _{z}/2}}{\Gamma \left (\alpha _{z}/2\right )}R^{\alpha _{R} +\alpha _{\varphi } -1}dR\vert \sin \varphi \vert ^{\alpha _{\varphi } -1}$ $\text{}\left \vert \text{}\cos \varphi \right \vert ^{\alpha _{R} -1}d\varphi \vert z\vert ^{\alpha _{z} -1}dz$. Using dimensionless cylindrical coordinates ($w_{R}^{ \prime }$, $\varphi ^{ \prime }$, $w_{z}^{ \prime }$), with $w_{R}^{ \prime } =R^{ \prime }/l_{0}$ and $w_{z}^{ \prime } =z^{ \prime }/l_{0}$, the volume integral of a function $f\left (w_{R}^{ \prime } ,\varphi ^{ \prime } ,w_{z}^{ \prime }\right )$ is then computed as:

\begin{equation}\int _{W}fd\mu _{H} =\frac{\pi ^{\left (\alpha _{R} +\alpha _{\varphi } +\alpha _{z}\right )/2}}{\Gamma \left (\alpha _{R}/2\right )\Gamma \left (\alpha _{\varphi }/2\right )\Gamma \left (\alpha _{z}/2\right )}\int w_{R}^{ \prime ^{\alpha _{R} +\alpha _{\varphi } -1}}dw_{R}^{ \prime }\int \left \vert \sin \varphi ^{ \prime }\right \vert ^{\alpha _{\varphi } -1}\left \vert \cos \varphi ^{ \prime }\right \vert ^{\alpha _{R} -1}d\varphi ^{ \prime }\int f(w_{R}^{ \prime } ,\varphi ^{ \prime } ,w_{z}^{ \prime })\left \vert w_{z}^{ \prime }\right \vert ^{\alpha _{z} -1}dw_{z}^{ \prime } , \label{eq3.9}
\end{equation}
where $0 <\alpha _{i} \leq 1$ for each dimension of the coordinate sub-spaces, and with the total dimension
$D =\alpha _{R} +\alpha _{\varphi } +\alpha _{z}$. 

For thin-disk structures, the NFDG potential is computed by combining together Eqs. (\ref{eq2.3}), (\ref{eq3.6}) (with rescaled coordinates), (\ref{eq3.9}), and the rescaled version of Eq. (\ref{eq3.2}), i.e.:
\begin{gather}\widetilde{\rho }\left (w_{R}^{ \prime } ,w_{z}^{ \prime }\right ) =\widetilde{\Sigma }\left (w_{R}^{ \prime }\right )\delta \left (w_{z}^{ \prime }\right ) \label{eq3.10} \\
\widetilde{\Sigma }\left (w_{R}^{ \prime }\right ) =\widetilde{\Sigma }_{0}e^{ -w_{R}^{ \prime }/W_{d}} =\frac{M}{2\pi W_{d}^{2}}e^{ -w_{R}^{ \prime }/W_{d}} . \nonumber \end{gather}

When performing the $w_{z}^{ \prime }$ integration in Eq. (\ref{eq3.9}), by using the Dirac delta function in Eq. (\ref{eq3.10}), we note that a finite result is obtained only for $\alpha _{z} =1$, i.e., no fractional dimension is needed in the $z^{ \prime }$ direction. There is instead some arbitrariness in the choice of the relation between $\alpha _{R}$ and $\alpha _{\varphi }$ and the overall space dimension $D$.  We will simply assume $\alpha _{R} =\alpha _{\varphi } =\alpha  =\frac{D -1}{2}$, so that the radial and angular coordinates will have the same (variable) fractional dimension.\protect\footnote{
We have also considered other possible choices for the $\alpha _{R}$, $\alpha _{z}$ values, such as having fractional dimension only in the radial direction ($\alpha _{R} =D -2$, $\alpha _{\varphi } =1$), or only in the angular direction ($\alpha _{R} =1$, $\alpha _{\varphi } =D -2$), for the thin-disk case ($\alpha _{z} =1$). The results do not differ much from those obtained with our preferred choice ($\alpha _{R} =\alpha _{\varphi } =\frac{D -1}{2}$), so we will not report them in this work. Also, results obtained with our preferred choice are somewhat in between those obtained with the other two extreme choices; thus, our choice for the $\alpha _{i}$ parameters can be considered a good average between all possible alternatives.
} In this case we also have: $D =2\alpha  +1 \leq 3$, and the results will depend only on the overall dimension $D$ of the space.

With these assumptions, we then obtain the potential in the $w_{z} =0$ plane as:

\begin{gather}\phi \left (w_{R}\right ) = -\frac{2\sqrt{\pi }\Gamma \left (D/2\right )G}{\left (D -2\right )\left [\Gamma \genfrac{(}{)}{}{}{D -1}{4}\right ]^{2}l_{0}^{2}}\sum \limits _{l =0}^{\infty }{\displaystyle\int _{0}^{\infty }}\widetilde{\Sigma }\left (w_{R}^{ \prime }\right )\frac{w_{R <}^{l}}{w_{R >}^{l +D -2}}w_{R}^{ \prime D -2}dw_{R}^{ \prime }{\displaystyle\int _{0}^{2\pi }}\left \vert \sin \varphi ^{ \prime }\right \vert ^{\frac{D -3}{2}}\left \vert \cos \varphi ^{ \prime }\right \vert ^{\frac{D -3}{2}}C_{l}^{\left (\frac{D}{2} -1\right )}\left (\cos \varphi ^{ \prime }\right )d\varphi ^{ \prime } \label{eq3.11} \\
= -\frac{\Gamma \left (D/2\right )GM}{\sqrt{\pi }\left (D -2\right )\left [\Gamma \genfrac{(}{)}{}{}{D -1}{4}\right ]^{2}l_{0}^{2}W_{d}^{2}}\sum \limits _{l =0}^{\infty }c_{l ,D}\left ({\displaystyle\int _{0}^{w_{R}}}e^{ -w_{R}^{ \prime }/W_{d}}\frac{w_{R}^{ \prime l}}{w_{R}^{l +D -2}}w_{R}^{ \prime D -2}dw_{R}^{ \prime } +{\displaystyle\int _{w_{R}}^{\infty }}e^{ -w_{R}^{ \prime }/W_{d}}\frac{w_{R}^{l}}{w^{ \prime \: l +D -2}_{R}}w_{R}^{ \prime D -2}dw_{R}^{ \prime }\right ) \nonumber  \\
= -\frac{\Gamma \left (D/2\right )GM}{\sqrt{\pi }\left (D -2\right )\left [\Gamma \genfrac{(}{)}{}{}{D -1}{4}\right ]^{2}l_{0}^{2}W_{d}^{2}}\sum \limits _{l =0}^{\infty }c_{l ,D}\left \{w_{R}\genfrac{(}{)}{}{}{W_{d}}{w_{R}}^{D +l -1}\left [\Gamma \left (D +l -1\right ) -\Gamma \left (D +l -1 ,\frac{w_{R}}{W_{d}}\right )\right ] +w_{R}E_{l}\genfrac{(}{)}{}{}{w_{R}}{W_{d}}\right \} . \nonumber \end{gather}

In the previous equation, $\Gamma \left (a ,z\right ) =\int _{z}^{\infty }t^{a -1}e^{ -t}dt$ is the incomplete gamma function, $E_{l}\left (z\right ) =\int _{1}^{\infty }e^{ -zt}t^{ -l}dt$ is the exponential integral function of order $l$, and we have denoted with constants $c_{l ,D}$ the results of the angular integrations for $D >1$, i.e.:

\begin{gather}c_{l ,D} ={\displaystyle\int _{0}^{2\pi }}\left \vert \sin \varphi ^{ \prime }\right \vert ^{\frac{D -3}{2}}\left \vert \cos \varphi ^{ \prime }\right \vert ^{\frac{D -3}{2}}C_{l}^{\left (\frac{D}{2} -1\right )}\left (\cos \varphi ^{ \prime }\right )d\varphi ^{ \prime } \label{eq3.12} \\
c_{0 ,D} = -\frac{2^{\frac{5 -D}{2}}\pi ^{3/2}\sec \genfrac{[}{]}{}{}{\pi \left (1 +D\right )}{4}}{\Gamma \genfrac{(}{)}{}{}{5 -D}{4}\Gamma \genfrac{(}{)}{}{}{1 +D}{4}} ;c_{2 ,D} =\frac{2^{\frac{1 -D}{2}}\pi ^{1/2}\left (D -2\right )^{2}\Gamma \genfrac{(}{)}{}{}{D -1}{4}}{\Gamma \genfrac{(}{)}{}{}{1 +D}{4}} ; . . . \nonumber  \\
c_{1 ,D} =c_{3 ,D} = . . . =0 \nonumber \end{gather}where these constants are identically zero for odd values of $l$, while they can be computed analytically for all even values of $l$.

The observed radial acceleration $\mathbf{g}_{obs}$, in the $w_{z} =0$ plane, can be obtained directly from Eq. (\ref{eq3.11}):

\begin{gather}\mathbf{g}_{obs}\left (w_{R}\right ) = -\frac{d\phi }{dw_{R}}\widehat{\mathbf{w}}_{R} =\frac{\Gamma \left (D/2\right )GM}{\sqrt{\pi }\left (D -2\right )\left [\Gamma \genfrac{(}{)}{}{}{D -1}{4}\right ]^{2}l_{0}^{2}W_{d}^{3}}\sum \limits _{l =0 ,2 ,4 , . . .}^{\infty }c_{l ,D}\Bigg\{ -w_{R}E_{l -1}\genfrac{(}{)}{}{}{w_{R}}{W_{d}} +W_{d}E_{l}\genfrac{(}{)}{}{}{w_{R}}{W_{d}} +\genfrac{(}{)}{}{}{W_{d}}{w_{R}}^{D +l}e^{ -w_{R}/W_{d}} \label{eq3.13} \\
 \times \left [W_{d}\genfrac{(}{)}{}{}{w_{R}}{W_{d}}^{D +l} -\left (D +l -2\right )w_{R}e^{ -w_{R}/W_{d}}\left (\Gamma \left (D +l -1\right ) -\Gamma \left (D +l -1 ,\frac{w_{R}}{W_{d}}\right )\right )\right ]\Bigg\}\widehat{\mathbf{w}}_{R} \nonumber \end{gather}
where only the terms for even values of $l$ need to be summed, while the terms for odd values of $l$ are identically zero, in view of Eq. (\ref{eq3.12}). We will use Eq. (\ref{eq3.13}) for $\mathbf{g}_{obs}$, as well as Eq. (\ref{eq3.4}) for $\mathbf{g}_{bar}$, in Sect. \ref{subsect:exponential} for the analysis of exponential thin-disk galaxies.

Another standard mass density distribution for thin-disk structures is the Kuzmin model, whose gravitational potential and surface mass density are, respectively \cite{2008gady.book.....B}:

\begin{gather}\phi _{K}\left (R ,z\right ) = -\frac{GM}{\sqrt{R^{2} +\left (R_{d} +\left \vert z\right \vert \right )^{2}}} \label{eq3.14} \\
\Sigma _{K}\left (R^{ \prime }\right ) =\Sigma _{0}\left (1 +\frac{R^{ \prime 2}}{R_{d}^{2}}\right )^{ -3/2} =\frac{M}{2\pi R_{d}^{2}}\left (1 +\frac{R^{ \prime 2}}{R_{d}^{2}}\right )^{ -3/2} , \nonumber \end{gather}
where the disk scale length is still denoted by $R_{d} >0$. Using rescaled coordinates, $w_{R} =R/l_{0}$ and $W_{d} =R_{d}/l_{0}$, we can rewrite the surface mass density and also obtain the gravitational field in the $z =0$ plane, directly from the potential in the first line of Eq. (\ref{eq3.14}):

\begin{gather}\widetilde{\Sigma }_{K}\left (w_{R}^{ \prime }\right ) =\widetilde{\Sigma }_{0}\left (1 +\frac{w_{R}^{ \prime 2}}{W_{d}^{2}}\right )^{ -3/2} =\frac{M}{2\pi W_{d}^{2}}\left (1 +\frac{w_{R}^{ \prime 2}}{W_{d}^{2}}\right )^{ -3/2} \label{eq3.15} \\
\mathbf{g}_{bar}\left (w_{R}\right ) = -\frac{GM}{l_{0}^{2}w_{R}^{2}}\left (1 +\frac{W_{d}^{2}}{w_{R}^{2}}\right )^{ -3/2}\widehat{\mathbf{w}}_{R} . \nonumber \end{gather}

The radial acceleration $\mathbf{g}_{obs}$ can be obtained with a procedure similar to the one outlined in Eqs. (\ref{eq3.5})-(\ref{eq3.13}) for the exponential disk. In particular, we obtain the potential in the $w_{z} =0$ plane as:

\begin{gather}\phi \left (w_{R}\right ) = -\frac{2\sqrt{\pi }\Gamma \left (D/2\right )G}{\left (D -2\right )\left [\Gamma \genfrac{(}{)}{}{}{D -1}{4}\right ]^{2}l_{0}^{2}}\sum \limits _{l =0}^{\infty }{\displaystyle\int _{0}^{\infty }}\widetilde{\Sigma }_{K}\left (w_{R}^{ \prime }\right )\frac{w_{R <}^{l}}{w_{R >}^{l +D -2}}w_{R}^{ \prime D -2}dw_{R}^{ \prime }{\displaystyle\int _{0}^{2\pi }}\left \vert \sin \varphi ^{ \prime }\right \vert ^{\frac{D -3}{2}}\left \vert \cos \varphi ^{ \prime }\right \vert ^{\frac{D -3}{2}}C_{l}^{\left (\frac{D}{2} -1\right )}\left (\cos \varphi ^{ \prime }\right )d\varphi ^{ \prime } \label{eq3.16} \\
= -\frac{\Gamma \left (D/2\right )GM}{\sqrt{\pi }\left (D -2\right )\left [\Gamma \genfrac{(}{)}{}{}{D -1}{4}\right ]^{2}l_{0}^{2}W_{d}^{2}}\sum \limits _{l =0}^{\infty }c_{l ,D}\left ({\displaystyle\int _{0}^{w_{R}}}\left (1 +\frac{w_{R}^{ \prime 2}}{W_{d}^{2}}\right )^{ -3/2}\frac{w_{R}^{ \prime l}}{w_{R}^{l +D -2}}w_{R}^{ \prime D -2}dw_{R}^{ \prime } +{\displaystyle\int _{w_{R}}^{\infty }}\left (1 +\frac{w_{R}^{ \prime 2}}{W_{d}^{2}}\right )^{ -3/2}\frac{w_{R}^{l}}{w^{ \prime \: l +D -2}_{R}}w_{R}^{ \prime D -2}dw_{R}^{ \prime }\right ) \nonumber  \\
= -\frac{\Gamma \left (D/2\right )GM}{\sqrt{\pi }\left (D -2\right )\left [\Gamma \genfrac{(}{)}{}{}{D -1}{4}\right ]^{2}l_{0}^{2}W_{d}^{2}}\sum \limits _{l =0}^{\infty }c_{l ,D}\Bigg\{\frac{w_{R}}{2(D +l -1)\left (w_{R}^{2} +W_{d}^{2}\right )}\bigg[2\left (D +l\right )W_{d}^{2}F\left ( -\frac{1}{2} ,\frac{1}{2}\left (D +l -1\right ) ;\frac{1}{2}\left (D +l +1\right ) ; -\frac{w_{R}^{2}}{W_{d}^{2}}\right ) \nonumber  \\
-2\left (\left (D +l -2\right )w_{R}^{2} +\left (D +l -1\right )W_{d}^{2}\right )F\left (\frac{1}{2} ,\frac{1}{2}\left (D +l -1\right ) ;\frac{1}{2}\left (D +l +1\right ) ; -\frac{w_{R}^{2}}{W_{d}^{2}}\right )\bigg] \nonumber  \\
+\frac{W_{d}^{3}}{\left (l +2\right )w_{R}^{2}\left (w_{R}^{2} +W_{d}^{2}\right )}\bigg[\left (l +3\right )w_{R}^{2}F\left ( -\frac{1}{2} ,\frac{l +2}{2} ;\frac{l +4}{2} ; -\frac{W_{d}^{2}}{w_{R}^{2}}\right ) -\left (\left (l +2\right )w_{R}^{2} +\left (l +1\right )W_{d}^{2}\right )F\left (\frac{1}{2} ,\frac{l +2}{2} ;\frac{l +4}{2} ; -\frac{W_{d}^{2}}{w_{R}^{2}}\right )\bigg]\Bigg\} . \nonumber \end{gather}

In the previous equation
we used the hypergeometric function defined by the Gauss series: $F\left (a ,b ;c ;z\right ) =\sum \limits _{s =0}^{\infty }\frac{(a)_{s}\left (b\right )_{s}}{\left (c\right )_{s}s !}z^{s}$,\protect\footnote{
The Pochhammer's symbol $\left (a\right )_{l}$ is defined as $\left (a\right )_{0} =1$, $\left (a\right )_{l} =a\left (a +1\right )\left (a +2\right ) . . .\left (a +l -1\right ) =\Gamma \left (a +l\right )/\Gamma \left (a\right )$.
} with the angular coefficients $c_{l ,D}$ computed as in Eq. (\ref{eq3.12}).

In the $w_{z} =0$ plane, the observed radial acceleration $\mathbf{g}_{obs}$ can then be obtained directly from Eq. (\ref{eq3.16}):

\begin{gather}\mathbf{g}_{obs}\left (w_{R}\right ) = -\frac{d\phi }{dw_{R}}\widehat{\mathbf{w}}_{R} =\frac{\Gamma \left (D/2\right )GM}{2\sqrt{\pi }\left (D -2\right )\left [\Gamma \genfrac{(}{)}{}{}{D -1}{4}\right ]^{2}l_{0}^{2}W_{d}^{2}\left (w_{R}^{2} +W_{d}^{2}\right )^{2}}\sum \limits _{l =0 ,2 ,4 , . . .}^{\infty }c_{l ,D}\Bigg\{\sqrt{1 +\frac{W_{d}^{2}}{w_{R}^{2}}}\left ( -4\frac{W_{d}^{5}}{w_{R}} -2W_{d}^{3}w_{R}\right ) \label{eq3.17} \\
 +\sqrt{1 +\frac{w_{R}^{2}}{W_{d}^{2}}}\left (2W_{d}^{4} +4W_{d}^{2}w_{R}^{2}\right ) +\frac{2(l +3)W_{d}^{3}}{\left (l +2\right )w_{R}}\left [\left (l +2\right )W_{d}^{2} +lw_{R}^{2}\right ]F\left ( -\frac{1}{2} ,\frac{l +2}{2} ;\frac{l +4}{2} ; -\frac{W_{d}^{2}}{w_{R}^{2}}\right ) \nonumber  \\
 -\frac{2(D +l)W_{d}^{2}}{\left (D +l -1\right )}\left [\left (D +l -2\right )W_{d}^{2} +\left (D +l\right )w_{R}^{2}\right ]F\left ( -\frac{1}{2} ,\frac{D +l -1}{2} ;\frac{D +l +1}{2} ; -\frac{w_{R}^{2}}{W_{d}^{2}}\right ) \nonumber  \\
 -\frac{2W_{d}^{3}}{\left (l +2\right )w_{R}^{3}}\left [\left (l +1\right )W_{d}^{2} +lw_{R}^{2}\right ]\left [lW_{d}^{2} +\left (l +2\right )w_{R}^{2}\right ]F\left (\frac{1}{2} ,\frac{l +2}{2} ;\frac{l +4}{2} ; -\frac{W_{d}^{2}}{w_{R}^{2}}\right ) \nonumber  \\
 +\frac{2}{\left (D +l -1\right )}[(D +l -2)\left (D +l -1\right )W_{d}^{4} +\left (8 +D\left (2D -7\right ) +l\left (4D -7\right ) +2l^{2}\right )W_{d}^{2}w_{R}^{2} \nonumber  \\
\left . +\left (D +l -2\right )^{2}w_{R}^{4}\right ]F\left (\frac{1}{2} ,\frac{D +l -1}{2} ;\frac{D +l +1}{2} ; -\frac{w_{R}^{2}}{W_{d}^{2}}\right )\Bigg\} , \nonumber \end{gather}
where, again, only the terms for even values of $l$ need to be summed (terms for odd values of $l$ are identically zero, in view of Eq. (\ref{eq3.12})). We will use Eq. (\ref{eq3.17}) for $\mathbf{g}_{obs}$, as well as Eq. (\ref{eq3.15}) for $\mathbf{g}_{bar}$, in Sect. \ref{subsect:Kuzmin} for the analysis of Kuzmin thin-disk models.

For thick-disk galaxies, it is customary \cite{2008gady.book.....B,Mannheim:2005bfa} to introduce a (rescaled)\  mass density:

\begin{equation}\widetilde{\rho }\left (w_{R}^{ \prime } ,w_{z}^{ \prime }\right ) =\widetilde{\Sigma }\left (w_{R}^{ \prime }\right )\widetilde{\zeta }\left (w_{z}^{ \prime }\right ) , \label{eq3.18}
\end{equation}where $\widetilde{\Sigma }$ can be the exponential function in Eq. (\ref{eq3.10}), the Kuzmin function in Eq. (\ref{eq3.15}), or others. The ``vertical'' density function $\widetilde{\zeta }(w_{z}^{ \prime })$ is usually chosen as one of the following \cite{Mannheim:2005bfa,2010ApJ...716..234B}:

\begin{gather}\widetilde{\zeta }_{1}\left (w_{z}^{ \prime }\right ) =\frac{1}{2H_{z}}e^{ -w_{z}^{ \prime }/H_{z}} \label{eq3.19} \\
\widetilde{\zeta }_{2}\left (w_{z}^{ \prime }\right ) =\frac{1}{4H_{z}}[\ensuremath{\operatorname*{sech}}\left (w_{z}^{ \prime }/2H_{z}\right )]^{2} \nonumber  \\
\widetilde{\zeta }_{3}\left (w_{z}^{ \prime }\right ) =\frac{1}{\sqrt{2}\pi H_{z}}\ensuremath{\operatorname*{sech}}\left (w_{z}^{ \prime }/\sqrt{2}H_{z}\right ) , \nonumber \end{gather}where we used rescaled versions of these structure functions \cite{2010ApJ...716..234B}, with the rescaled parameter $H_{z} =h_{z}/l_{0}$ connected with the original vertical scale height $h_{z}$.

All these functions are symmetric, i.e., $\widetilde{\zeta }\left ( -w_{z}^{ \prime }\right ) =\widetilde{\zeta }\left (w_{z}^{ \prime }\right )$ and normalized ($\int _{ -\infty }^{\infty }\widetilde{\zeta }\left (w_{z}^{ \prime }\right )dw_{z}^{ \prime } =1$), thus recovering the thin-disk $\delta \left (w_{z}^{ \prime }\right )$ in Eq. (\ref{eq3.10}) for $H_{z} \rightarrow 0$. We will also adopt the standard relation \cite{Lelli:2016zqa,2010ApJ...716..234B}, $\left (h_{z}/\ensuremath{\operatorname*{kpc}}\right ) =0.196\left (R_{d}/\ensuremath{\operatorname*{kpc}}\right )^{0.633}$, between the vertical scale height $h_{z}$ and the radial scale length $R_{d}$ in Eq. (\ref{eq3.2}), properly rescaled by using our dimensionless variables.

Therefore, for thick-disk structures we obtain the potential $\phi \left (w_{R}\right )$ in the $w_{z} =0$ plane by following similar steps used for the thin-disk case above. For simplicity, we will just choose the exponential vertical function $\widetilde{\zeta }_{1}$ in Eq. (\ref{eq3.19}), together with the exponential radial mass density $\widetilde{\Sigma }$ from Eq. (\ref{eq3.10}), and the connection between their respective scale lengths described in the previous paragraph. We will enter these functions into the general integral for $\phi $ in Eqs. (\ref{eq2.3}) and (\ref{eq3.9}), using the expansion for the Euler kernel in Eq. (\ref{eq3.5}) supplemented with the coordinate transformations in Eq. (\ref{eq3.7}).

We will also keep the same choice, for the $\alpha _{R}$, $\alpha _{\varphi }$, $\alpha _{z}$  parameters, which was used in the thin-disk case: $\alpha _{R} =\alpha _{\varphi } =\frac{D -1}{2}$ and $\alpha _{z} =1$, with $D =D\left (w_{R}\right ) .$ The choice for $\alpha _{R}$ and $\alpha _{\varphi }$ follows the idea of sharing equally the fractional dimension in both radial and angular directions. The choice of simply assuming $\alpha _{z} =1$ is due to the fact that in our model the dimension is a function of the field point. Since our observations are done in the $z =0$ plane, the value of $\alpha _{z}\left (z =0\right ) =1$ follows by assuming Newtonian behaviour at $z =0$ (or simply by continuity with respect to the thin-disk case, where we also assumed $\alpha _{z} =1$).

In this way, the gravitational potential $\phi \left (w_{R}\right )$, for $1 <D \leq 3$, can be determined through a triple numerical integration of the different terms in the general series obtained from the kernel expansion. The observed radial acceleration $\mathbf{g}_{obs}$, in the $w_{z} =0$ plane, can then be obtained by a simple radial derivative of the potential. Since for the thick-disk case we use the Euler kernel in Eq. (\ref{eq3.5}) together with the coordinate transformations in Eq. (\ref{eq3.7}), the angular integrals cannot be separated from the radial and vertical integrals and the computation has to be carried out entirely in a numerical way.\protect\footnote{
All the numerical computations (and some of the analytical ones) in this work, were performed with Mathematica, Version 12.1.1.0, Wolfram Research Inc.
}

The resulting radial acceleration $\mathbf{g}_{obs}$ can be compared with the standard baryonic $\mathbf{g}_{bar}$, obtained with the same procedure, but with a fixed $D =3$ value, or with equivalent methods for thick-disks in the literature \cite{Mannheim:2005bfa}. In Sect. \ref{subsect:exponentialthick}, we will show results of these computations using the techniques outlined above.

Finally, we want to remark that the general procedure outlined in this section can also be adapted to spherically symmetric structures, such as galactic spherical bulges, globular clusters, or others. In fact, the NFDG potential expansion in Eq. (\ref{eq3.5}) is better suited to spherical coordinates, rather than cylindrical. Aligning the field vector $\mathbf{r}$ in the direction of the $z^{ \prime }$ axis and using standard spherical coordinates ($r^{ \prime }$, $\theta ^{ \prime }$, $\varphi ^{ \prime }$) for the source vector $\mathbf{r}^{ \prime }$, we can immediately use expansion (\ref{eq3.5}) with the angle $\gamma $ replaced by $\theta ^{ \prime }$.

In order to compute the gravitational potential $\phi \left (w_{r}\right )$ for a spherically symmetric mass distribution ($w_{r} =r/l_{0}$), described by a mass density $\widetilde{\rho }\left (w_{r}^{ \prime }\right )$, we can still use our main Eq. (\ref{eq2.3}), but the volume integration must be performed in spherical coordinates. As described in our paper I, using dimensionless spherical coordinates ($w_{r}^{ \prime }$, $\theta ^{ \prime }$, $\varphi ^{ \prime }$) with $w_{r}^{ \prime } =r^{ \prime }/l_{0}$, the volume integral of a function $f\left (w_{r}^{ \prime } ,\theta ^{ \prime } ,\varphi ^{ \prime }\right )$ can be computed as:

\begin{gather}\int _{W}fd\mu _{H} =\frac{\pi ^{\left (\alpha _{r} +\alpha _{\theta } +\alpha _{\varphi }\right )/2}}{\Gamma \left (\alpha _{r}/2\right )\Gamma \left (\alpha _{\theta }/2\right )\Gamma \left (\alpha _{\varphi }/2\right )}\int w_{r}^{ \prime \: \alpha _{r} +\alpha _{\theta } +\alpha _{\varphi } -1}dw_{r}^{ \prime } \label{eq3.20} \\
 \times \int \left \vert \sin \theta ^{ \prime }\right \vert ^{\alpha _{r} +\alpha _{\theta } -1}\left \vert \cos \theta ^{ \prime }\right \vert ^{\alpha _{\varphi } -1}d\theta ^{ \prime }\int f(w_{r}^{ \prime } ,\theta ^{ \prime } ,\varphi ^{ \prime })\left \vert \sin \varphi ^{ \prime }\right \vert ^{\alpha _{\theta } -1}\left \vert \cos \varphi ^{ \prime }\right \vert ^{\alpha _{r} -1}d\varphi ^{ \prime } , \nonumber \end{gather}
where $0 <\alpha _{i} \leq 1$ for each dimension of the coordinate sub-spaces, and with the total dimension
$D =\alpha _{r} +\alpha _{\theta } +\alpha _{\varphi }$.

Combining Eqs. (\ref{eq2.3}), (\ref{eq3.20}) with $f\left (w_{r}^{ \prime } ,\theta ^{ \prime } ,\varphi ^{ \prime }\right ) =\widetilde{\rho }\left (w_{r}^{ \prime }\right )$, and expansion (\ref{eq3.5}) with the angle $\gamma $ replaced by $\theta ^{ \prime }$, the volume integral is completely separable with the following angular integrals:

\begin{gather}c_{l ,\alpha _{r} ,\alpha _{\theta } ,\alpha _{\varphi }} =\int _{0}^{\pi }\left \vert \sin \theta ^{ \prime }\right \vert ^{\alpha _{r} +\alpha _{\theta } -1}\left \vert \cos \theta ^{ \prime }\right \vert ^{\alpha _{\varphi } -1}C_{l}^{\left (\frac{D}{2} -1\right )}\left (\cos \theta ^{ \prime }\right )d\theta ^{ \prime } \label{eq3.21} \\
c_{0 ,\alpha _{r} ,\alpha _{\theta } ,\alpha _{\varphi }} =\frac{\pi \csc \genfrac{(}{)}{}{}{\pi \alpha _{\varphi }}{2}\Gamma \genfrac{(}{)}{}{}{\alpha _{r} +\alpha _{\theta }}{2}}{\Gamma \left (1 -\frac{\alpha _{\varphi }}{2}\right )\Gamma \genfrac{(}{)}{}{}{D}{2}} ;c_{2 ,\alpha _{r} ,\alpha _{\theta } ,\alpha _{\varphi }} =\frac{\pi \left (\alpha _{\varphi } -1\right )\csc \genfrac{(}{)}{}{}{\pi \alpha _{\varphi }}{2}\Gamma \left (\frac{\alpha _{r} +\alpha _{\theta }}{2}\right )}{\Gamma \left (1 -\frac{\alpha _{\varphi }}{2}\right )\Gamma \left (\frac{D}{2} -1\right )} ; . . . \nonumber  \\
\int _{0}^{2\pi }\left \vert \sin \varphi ^{ \prime }\right \vert ^{\alpha _{\theta } -1}\left \vert \cos \varphi ^{ \prime }\right \vert ^{\alpha _{r} -1}d\varphi ^{ \prime } =\frac{2\pi \csc \genfrac{(}{)}{}{}{\pi \alpha _{r}}{2}\Gamma \genfrac{(}{)}{}{}{\alpha _{\theta }}{2}}{\Gamma \left (1 -\frac{\alpha _{r}}{2}\right )\Gamma \genfrac{(}{)}{}{}{\alpha _{r} +\alpha _{\theta }}{2}} , \nonumber \end{gather}where we have denoted the results of the angular integrals in the first line of the last equation with constants $c_{l ,\alpha _{r} ,\alpha _{\theta } ,\alpha _{\varphi }}$. These $\theta ^{ \prime }$ integrals can be computed analytically for all values of $l =0 ,1 ,2 , . . .$ They are all identically zero for odd values of $l$, as was the case of the similar constants in Eq. (\ref{eq3.12}).

As in the previous cases, we will assume that the fractional dimension applies to all coordinates equally, i.e., $\alpha _{r} =\alpha _{\theta } =\alpha _{\varphi } =D/3$, so that the angular integrals in Eq. (\ref{eq3.21}) simplify as:
\begin{gather}c_{l ,D} =\int _{0}^{\pi }\left \vert \sin \theta ^{ \prime }\right \vert ^{\frac{2D}{3} -1}\left \vert \cos \theta ^{ \prime }\right \vert ^{\frac{D}{3} -1}C_{l}^{\left (\frac{D}{2} -1\right )}\left (\cos \theta ^{ \prime }\right )d\theta ^{ \prime } \label{eq3.23} \\
c_{0 ,D} =\frac{\pi \csc \genfrac{(}{)}{}{}{\pi D}{6}\Gamma \genfrac{(}{)}{}{}{D}{3}}{\Gamma \left (1 -\frac{D}{6}\right )\Gamma \genfrac{(}{)}{}{}{D}{2}} ;c_{2 ,D} =\frac{\pi \left (\frac{D}{3} -1\right )\csc \genfrac{(}{)}{}{}{\pi D}{6}\Gamma \left (\frac{D}{3}\right )}{\Gamma \left (1 -\frac{D}{6}\right )\Gamma \left (\frac{D}{2} -1\right )} ; . . . \nonumber  \\
\int _{0}^{2\pi }\left \vert \sin \varphi ^{ \prime }\right \vert ^{\frac{D}{3} -1}\left \vert \cos \varphi ^{ \prime }\right \vert ^{\frac{D}{3} -1}d\varphi ^{ \prime } =\frac{2\pi \csc \genfrac{(}{)}{}{}{\pi D}{6}\Gamma \genfrac{(}{)}{}{}{D}{6}}{\Gamma \left (1 -\frac{D}{6}\right )\Gamma \genfrac{(}{)}{}{}{D}{3}} . \nonumber \end{gather}

Using the previous equations and after further simplifications, the general potential for spherically symmetric mass distributions can be written as:

\begin{gather}\phi \left (w_{r}\right ) = -\frac{2\pi \Gamma \left (\frac{D}{2} -1\right )G}{\Gamma \left (\frac{D}{3}\right )\Gamma \genfrac{(}{)}{}{}{D}{6}l_{0}^{2}}\sum \limits _{l =0 ,2 ,4 , . . .}^{\infty }c_{l ,D}{\displaystyle\int _{0}^{\infty }}\widetilde{\rho }\left (w_{r}^{ \prime }\right )\frac{w_{r <}^{l}}{w_{r >}^{l +D -2}}w_{r}^{ \prime D -1}dw_{r}^{ \prime } \label{eq3.24} \\
 = -\frac{2\pi \Gamma \left (\frac{D}{2} -1\right )G}{\Gamma \left (\frac{D}{3}\right )\Gamma \genfrac{(}{)}{}{}{D}{6}l_{0}^{2}}\sum \limits _{l =0 ,2 ,4 , . . .}^{\infty }c_{l ,D}\left ({\displaystyle\int _{0}^{w_{r}}}\widetilde{\rho }\left (w_{r}^{ \prime }\right )\frac{w_{r}^{ \prime l}}{w_{r}^{l +D -2}}w_{r}^{ \prime D -1}dw_{r}^{ \prime } +{\displaystyle\int _{w_{r}}^{\infty }}\widetilde{\rho }\left (w_{r}^{ \prime }\right )\frac{w_{r}^{l}}{w^{ \prime \: l +D -2}_{r}}w_{r}^{ \prime D -1}dw_{r}^{ \prime }\right ) \nonumber \end{gather}and the related observed radial acceleration $\mathbf{g}_{obs}\left (w_{r}\right )$ can be obtained directly by differentiation of the last equation. The standard ($D =3$) radial acceleration $\mathbf{g}_{bar}$ can be simply obtained from Eq. (\ref{eq2.12}). In Sect. \ref{subsect:spherical} we will apply this method to a simple Plummer spherical model and compare the results with those obtained in paper I.

As a final consideration, we note that all our formulas for the gravitational potential $\phi $ and for the NFDG gravitational field $\mathbf{g}_{obs}$, such as those in Eqs. (\ref{eq3.11}), (\ref{eq3.13}), (\ref{eq3.16}), (\ref{eq3.17}), and (\ref{eq3.24}), require summing all non-zero terms for $l =0 ,2 ,4 , . . .$ (terms for odd values of $l$ being identically zero). Practically, we found that these series of functions converge rather quickly over the whole range of $w_{R} >0$ (or $w_{r} >0$), with just the exception of the thick-disk formulas for very low values of $w_{R}$.

All the results presented in the following sections were computed by summing the first few terms (typically the first six non-zero terms, for $l =0 ,2 ,4 ,6 ,8 ,10$) of our NFDG expansions. For each case, we also tested numerically that our NFDG\ expansions correctly reduce, for fixed $D =3$, to the standard $\mathbf{g}_{bar}$ expressions in the literature, again by summing only the first few non-zero terms of our ($D =3$) NFDG\ expansions. Therefore, we are confident that our formulas are mathematically sound and can describe accurately the physical reality of galactic structures in spaces of dimension $D \leq 3$.

\section{\label{sect::galactic} Galactic models
}
From the analysis presented in the previous sections and in paper I, our NFDG has been introduced as a modification of the law of gravity and
not of the law of inertia and Newtonian dynamics. As in our previous work \cite{Varieschi:2020ioh}, we will assume that Newtonian dynamics is not affected in any way by our
fractional generalizations. A test object, subject to a fractional gravitational field, such as those described by Eqs. (\ref{eq3.13}) and (\ref{eq3.17}), will still move in a
(classical)
$3 +1$
space-time, thus obeying standard laws of dynamics.

 In the
following sub-sections, we will apply NFDG to some fundamental galactic structures, particularly to axially-symmetric cases,  and connect
our
results with the empirical MOND predictions outlined in Sect. \ref{sect:REVIEW}.

\subsection{\label{subsect:exponential}
Exponential thin disks
}

 We will start with the case of a thin-disk galaxy, with an exponential surface mass density described in Eq. (\ref{eq3.10}). Following our NFDG assumptions, we define
$(g_{obs}/g_{bar})_{NFDG}$
as the ratio of the magnitude of $\mathbf{g}_{obs}$ from Eq. (\ref{eq3.13}) and the magnitude of $\mathbf{g}_{bar}$ from Eq. (\ref{eq3.4}):

\begin{equation}\genfrac{(}{)}{}{}{g_{obs}}{g_{bar}}_{NFDG}(w_{R}) =\frac{\left \vert \mathbf{g}_{obs}\left (w_{R} ,D\left (w_{R}\right )\right )\right \vert }{\left \vert \mathbf{g}_{bar}\left (w_{R}\right )\right \vert }^{\left .\right .} , \label{eq4.1}
\end{equation}
where the gravitational field from Eq. (\ref{eq3.13}) is now denoted as $\mathbf{g}_{obs}\left (w_{R} ,D\left (w_{R}\right )\right )$ to signify that the dimension $D$ in the same equation should now be considered a function of the field point radial coordinate, i.e., $D =D(w_{R})$, but not a function of the angular coordinate $\varphi $, due to the axial symmetry.
This function needs to be determined either from experimental data or from theoretical considerations.
In this sub-section and in the following ones, we will consider just the first option, while a possible theoretical determination of the
dimension function will discussed in Sect. \ref{subsect:discussion}, for the case of the NGC 6503 galaxy.

It should be noted again that the NFDG main equations (\ref{eq2.2})-(\ref{eq2.3}), and the other fundamental equations introduced in paper I, were derived for a fixed value of the dimension $D$, while in this section we assume a variable dimension $D(w_{R})$. In the same paper I, we justified this transition from constant to variable dimension by assuming a slow change over galactic distances of this variable dimension, so that the fundamental NFDG equations are still approximately valid when replacing $D$ with $D(w_{R})$. 
However, this transition from constant to variable dimension should be introduced in a more rigorous way, following the existing studies in the literature \cite{Calcagni:2011sz,Calcagni:2016azd,Calcagni:2018dhp,Carlip:2019onx,Calcagni:2016xtk}. It is beyond the scope of this paper to include this analysis in the current work, but the issue will be addressed in an upcoming publication \cite{Varieschi:2020hvp}.

To obtain
$D(w_{R})$
from experimental data, without fitting any particular set of galactic data, we will follow the same procedure used in paper I: compare the expression in Eq. (\ref{eq4.1}) with the MOND
equivalent expression in Eq. (\ref{eq2.8}),
$\genfrac{(}{)}{}{}{g_{obs}}{g_{bar}}_{MOND}\left (w_{R}\right ) =\frac{1}{1 -e^{ -\sqrt{g_{bar}\left (w_{R}\right )/g_{\dag }}}}$, or with similar expressions obtained by using the other interpolation functions in Eq.
(\ref{eq2.7}), and solve numerically the resulting equation for the dimension $D$ at each field point $w_{R}$.  

If the two expressions for $\genfrac{(}{)}{}{}{g_{obs}}{g_{bar}}$ are compatible, we expect to obtain
$D(w_{R}^{})$
as a continuous function with values
$D \approx 3$
in regions where Newtonian gravity holds. The dimension should then decrease
toward $D \approx 2$ in regions where the DML applies, following our general discussion in Sect. \ref{sect:REVIEW}.

\begin{figure}\centering 
\setlength\fboxrule{0in}\setlength\fboxsep{0.1in}\fcolorbox[HTML]{FFFFFF}{FFFFFF}{\includegraphics[ width=7.03in, height=6.789414153132252in,]{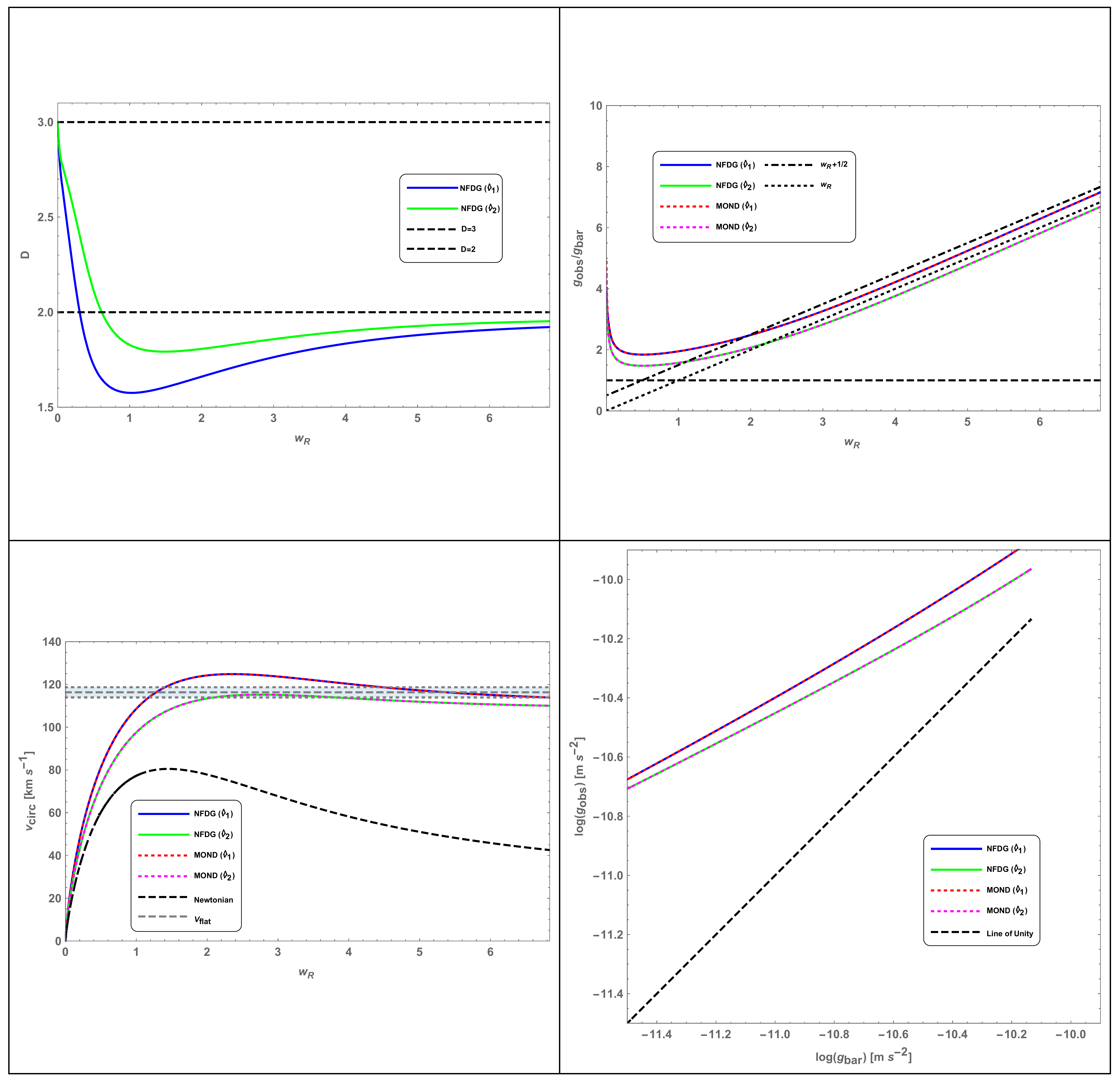}
}
\caption{Thin-disk exponential galaxy results.
Top-left panel: NFDG variable dimension $D\left (w_{R}\right )$ for MOND interpolation functions $\widehat{\nu }_{1}$ and $\widehat{\nu }_{2}$. Other panels: comparison of NFDG\ results (solid lines) with equivalent MOND predictions (dotted lines) for the two different interpolation functions. Also shown: Newtonian behavior-Line of Unity (black-dashed lines).
}\label{figure:thindiskexponential}\end{figure}

 Figure \ref{figure:thindiskexponential} shows all
the results for this particular case. The top-left panel illustrates the variable dimension
$D\left (w_{R}\right )$
obtained using functions $\widehat{\nu }_{1}$ and $\widehat{\nu }_{2}$ in Eq. (\ref{eq2.7}). The dimension functions are uniquely
defined and continuous over the whole range: at low-$w_{R}$,
$D \approx 3$
in the Newtonian regime,\protect\footnote{
In this figure, as well as in the other similar figures of this section, we assumed $D =3$ at the origin $w_{R} =0$.
} then the dimension decreases toward a minimum, and eventually approaches the value
$D \approx 2$
in the deep-MOND regime, as expected.

 The top-right panel in the figure shows the ratio
$\genfrac{(}{)}{}{}{g_{obs}}{g_{bar}}$
computed in two different ways:
$\genfrac{(}{)}{}{}{g_{obs}}{g_{bar}}_{NFDG}\left (w_{R}\right )$
from Eq. (\ref{eq4.1}) with the dimension functions
$D\left (w_{R}\right )$
obtained before, and
$\genfrac{(}{)}{}{}{g_{obs}}{g_{bar}}_{MOND}\left (w_{R}\right ) =\widehat{\nu }_{1}\left (w_{R}\right )$
(or
$\widehat{\nu }_{2}\left (w_{R}\right )$), simply using the two MOND functions in Eq. (\ref{eq2.7}). In both cases, the NFDG plots (solid lines) match the MOND ones (dotted lines). At low
$w_{R}$,
within the Newtonian regime, we don't simply have
$\frac{g_{obs}}{g_{bar}} \simeq 1$, because the RAR ratio $\genfrac{(}{)}{}{}{g_{obs}}{g_{bar}}_{MOND}\left (w_{R}\right )$ diverges for $w_{R} \rightarrow 0$ (as $g_{bar} \rightarrow 0$) and the NFDG\ ratio also follows this divergence for $w_{R} \rightarrow 0$. In the deep-MOND high-$w_{R}$
range we have instead
$\frac{g_{obs}}{g_{bar}} \sim w_{R} +\frac{1}{2}$, or $\frac{g_{obs}}{g_{bar}} \sim w_{R}$, for the two cases related to $\widehat{\nu }_{1}$ and $\widehat{\nu }_{2}$ respectively, as expected in the MOND model. 

The results shown in these two top panels of figure \ref{figure:thindiskexponential} are independent of the total mass
$M$
of the thin-disk object, and were obtained by using only the
$n =1 ,2$
values for the general MOND function $\widehat{\nu }_{n}$ in Eq. (\ref{eq2.7}). Using
$n >2$
values for the same function $\widehat{\nu }_{n}$ does not yield results which are much different from the
$n =2$
ones.\protect\footnote{ Using MOND functions
$\nu _{n}$,
instead of
$\widehat{\nu }_{n}$,
yields very similar results for all values of $n$. Therefore, we have considered only the $\widehat{\nu }_{n}$ family of MOND interpolation functions as was also done in our previous paper I.}

 The bottom-left panel shows circular velocity plots corresponding to the
previously analyzed cases, and compared with the purely Newtonian case. For this panel, as well as for the
bottom-right one, we have assumed a total mass $M =1.72 \times 10^{40}\thinspace \mbox{kg}$, with
$l_{0} \approx \sqrt{\frac{GM}{a_{0}}} \simeq 9.79 \times 10^{19}\mbox{m}$, and disk scale length $R_{d} =2.16\ \ensuremath{\operatorname*{kpc}} =6.67 \times 10^{19}\ \mbox{m}$, (rescaled length $W_{d} =R_{d}/l_{0} =0.681$). These values refer to the field dwarf spiral galaxy NGC 6503 \cite{Lelli:2016zqa}, which will be studied in detail in Sect. \ref{sect:6503}. We will use these reference values for most of the cases studied in this paper, but the results presented are largely independent of the choice of mass $M$, or other physical parameters.

In this bottom-left panel the NFDG circular speeds are
computed as
$v_{circ} =\sqrt{g_{obs}\left (w_{R}\right )w_{R}\thinspace l_{0}}/10^{3}$$\left [\mbox{km}\thinspace \mbox{s}^{ -1}\right ]$,
while the MOND circular speeds are computed as
$v_{circ} =\sqrt{g_{bar}\left (w_{R}\right )\widehat{\nu }_{1}\left (w_{R}\right )w_{R}\thinspace l_{0}}/10^{3}\left [\mbox{km}\thinspace \mbox{s}^{ -1}\right ]$
(or using
$\widehat{\nu }_{2}$
instead of
$\widehat{\nu }_{1}$), and the purely Newtonian speed is
$v_{circ} =\sqrt{g_{bar}\left (w_{R}\right )w_{R}\thinspace l_{0}}/10^{3}\left [\mbox{km}\thinspace \mbox{s}^{ -1}\right ]$. As seen from the panel, there is perfect agreement between the respective ($\widehat{\nu }_{1}$
or
$\widehat{\nu }_{2}$) NFDG and MOND cases, showing the expected flattening of the circular speed plots at
high-$w_{R}$, as opposed to the standard Newtonian decrease of circular speed with radial distance. In the same panel, we also show the NGC 6503 flat rotation velocity
$V_{f} =116.3 \pm 2.4\left [\mbox{km}\thinspace \mbox{s}^{ -1}\right ]$  \cite{Lelli:2016zqa}, represented by the horizontal gray lines and gray band in the figure. Both NFDG and MOND $\left (\widehat{\nu }_{1}\right )$ velocity curves are in agreement with this $V_{f}$ value at
high-$w_{R}$ as expected, while the NFDG and MOND $\left (\widehat{\nu }_{2}\right )$ curves are a bit less consistent with the $V_{f}$ value.

In these first three panels of Fig. \ref{figure:thindiskexponential}, we plotted all our results in terms of the rescaled cylindrical radial coordinate $w_{R}$, rather than the physical radial coordinate $R$, since the general behavior of the NFDG analysis is independent of the actual physical parameters of the thin-disk galaxy being considered. We also plotted all our results up to a maximum value of $w_{R}$ given by $w_{R ,\max } \approx 10\ W_{d}$ since about  $99.9 \%$ of the galaxy mass is contained within this limiting value, following the exponential distribution in Eq. (\ref{eq3.10}).

Finally, the bottom-right panel is similar to the
$\log \left (g_{obs}\right )$
vs.
$\log \left (g_{bar}\right )$
plots widely used in the literature (see Fig. 3 in Ref. \cite{McGaugh:2016leg} or the figures in Ref. \cite{Lelli:2017vgz}) to illustrate the validity of the general MOND-RAR relation from Eq.
(\ref{eq2.8}). Compared to the \textit{Line of Unity},
representing the purely Newtonian case, there is agreement between plots obtained with our NFDG\ model, using
$g_{obs}\left (w_{R}\right )$
and
$g_{bar}\left (w_{R}\right )$
from Eq. (\ref{eq3.13}) and (\ref{eq3.4}) respectively, and MOND plots where
$g_{obs}\left (w_{R}\right ) =\widehat{\nu }_{1}\left (w_{R}\right )g_{bar}\left (w_{R}\right )$
(or
$g_{obs}\left (w_{R}\right ) =\widehat{\nu }_{2}\left (w_{R}\right )g_{bar}\left (w_{R}\right )$).

 This study of the thin-disk galaxy case already shows that the variable-dimension effect of NFDG
can be equivalent to the MOND-RAR model. In the next three sub-sections we will confirm this result
using other axially/spherically symmetric cases.

\subsection{\label{subsect:Kuzmin}
Kuzmin thin disks
}

 As our second case for a thin-disk galaxy, we consider a Kuzmin model with surface density descibed by Eq. (\ref{eq3.14}), or by the rescaled Eq. (\ref{eq3.15}) which also includes the simple expression for
$\mathbf{g}_{bar}\left (w_{R}\right )$. The NFDG potential $\phi \left (w_{R}\right )$ in the $z =0$ plane and the related field
$\mathbf{g}_{obs}\left (w_{R}\right )$ are described by Eqs.(\ref{eq3.16})-(\ref{eq3.17}), so that the ratio of the field magnitudes
$\genfrac{(}{)}{}{}{g_{obs}}{g_{bar}}_{NFDG}\left (w_{R}\right ) =\frac{\left \vert \mathbf{g}_{obs}\left (w_{R} ,D\left (w_{R}\right )\right )\right \vert }{\left \vert \mathbf{g}_{bar}\left (w_{R}\right )\right \vert }$ can be easily computed analytically.

As in the previous exponential case, the dimension
$D\left (w_{R}\right )$
is obtained by solving numerically the equation
$\genfrac{(}{)}{}{}{g_{obs}}{g_{bar}}_{NFDG}\left (w_{R}\right ) =\widehat{\nu }_{1}\left (w_{R}\right )$
(or
$\widehat{\nu }_{2}\left (w_{R}\right )$). The total mass $M$, the length scale $R_{d}$, and the other parameters are chosen to be the same as those used in Sect. \ref{subsect:exponential}.

\begin{figure}\centering 
\setlength\fboxrule{0in}\setlength\fboxsep{0.1in}\fcolorbox[HTML]{FFFFFF}{FFFFFF}{\includegraphics[ width=7.08in, height=6.835862068965518in,]{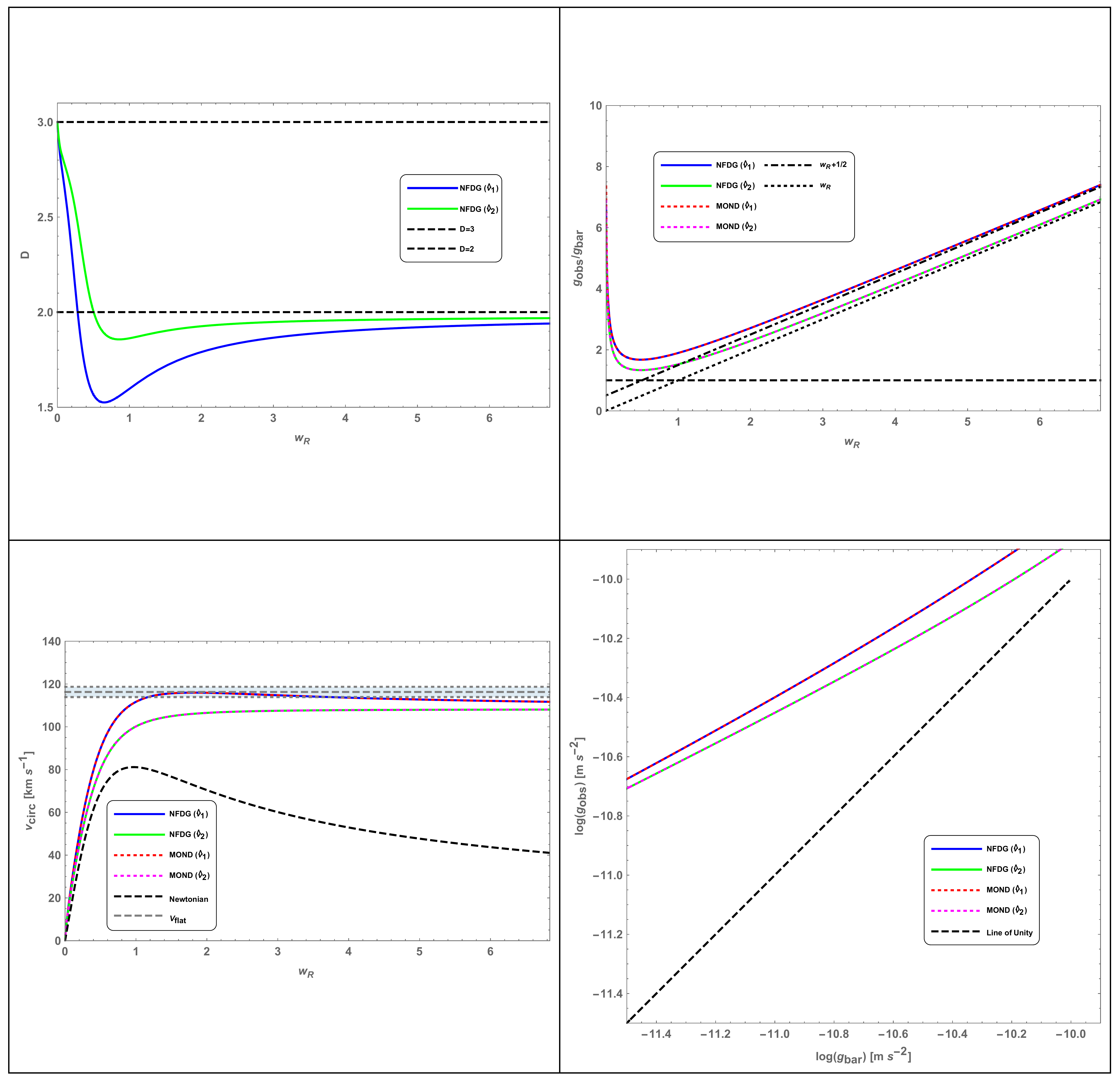}
}
\caption{Thin-disk Kuzmin galaxy results.
Top-left panel: NFDG variable dimension $D\left (w_{R}\right )$ for MOND interpolation functions $\widehat{\nu }_{1}$ and $\widehat{\nu }_{2}$. Other panels: comparison of NFDG\ results (solid lines) with equivalent MOND predictions (dotted lines) for the two different interpolation functions. Also shown: Newtonian behavior-Line of Unity (black-dashed lines).
}\label{figure:kuzmin}\end{figure}

 Figure \ref{figure:kuzmin}
shows the results for this case, in the same way of Fig. \ref{figure:thindiskexponential}, previously. The top left panel illustrates the dimension functions
$D\left (w_{R}\right )$
for the two cases being considered. Again, the dimension functions are uniquely
defined and continuous over the whole range: at low-$w_{R}$
values,
$D \approx 3$
in the Newtonian regime, then the dimension eventually approaches the value
$D \approx 2$
in the DML, as expected.

 The top-right panel shows the same two regimes, Newtonian and deep-MOND, in terms
of the
$\genfrac{(}{)}{}{}{g_{obs}}{g_{bar}}$
ratio: closer to unity at low-$w_{R}$ (Newtonian), but still diverging for $w_{R} \rightarrow 0$, and approaching asymptotically
$w_{R} +1/2$
(or $w_{R}$) at high-$w_{R}$
(deep-MOND). Finally, the two bottom panels show the equivalent circular speeds and
log-log plots, with perfect correspondence between the NFDG and MOND computations (obtained with the
same procedure outlined above for figure \ref{figure:thindiskexponential}). This time, the Kuzmin model is less successful in recovering the NGC 6503 flat rotation velocity
$V_{f} \simeq 116.3\ \mbox{km}\thinspace \mbox{s}^{ -1}$, shown in the bottom-left panel of the figure. Exponential models are usually more effective in the case of thin-disk galaxies, but we wanted to include also the Kuzmin model in our analysis as a second example of a fully analytical computation in NFDG.

\subsection{\label{subsect:exponentialthick}Exponential thick disks
}
\begin{figure}\centering 
\setlength\fboxrule{0in}\setlength\fboxsep{0.1in}\fcolorbox[HTML]{FFFFFF}{FFFFFF}{\includegraphics[ width=7.08in, height=6.836955223880597in,]{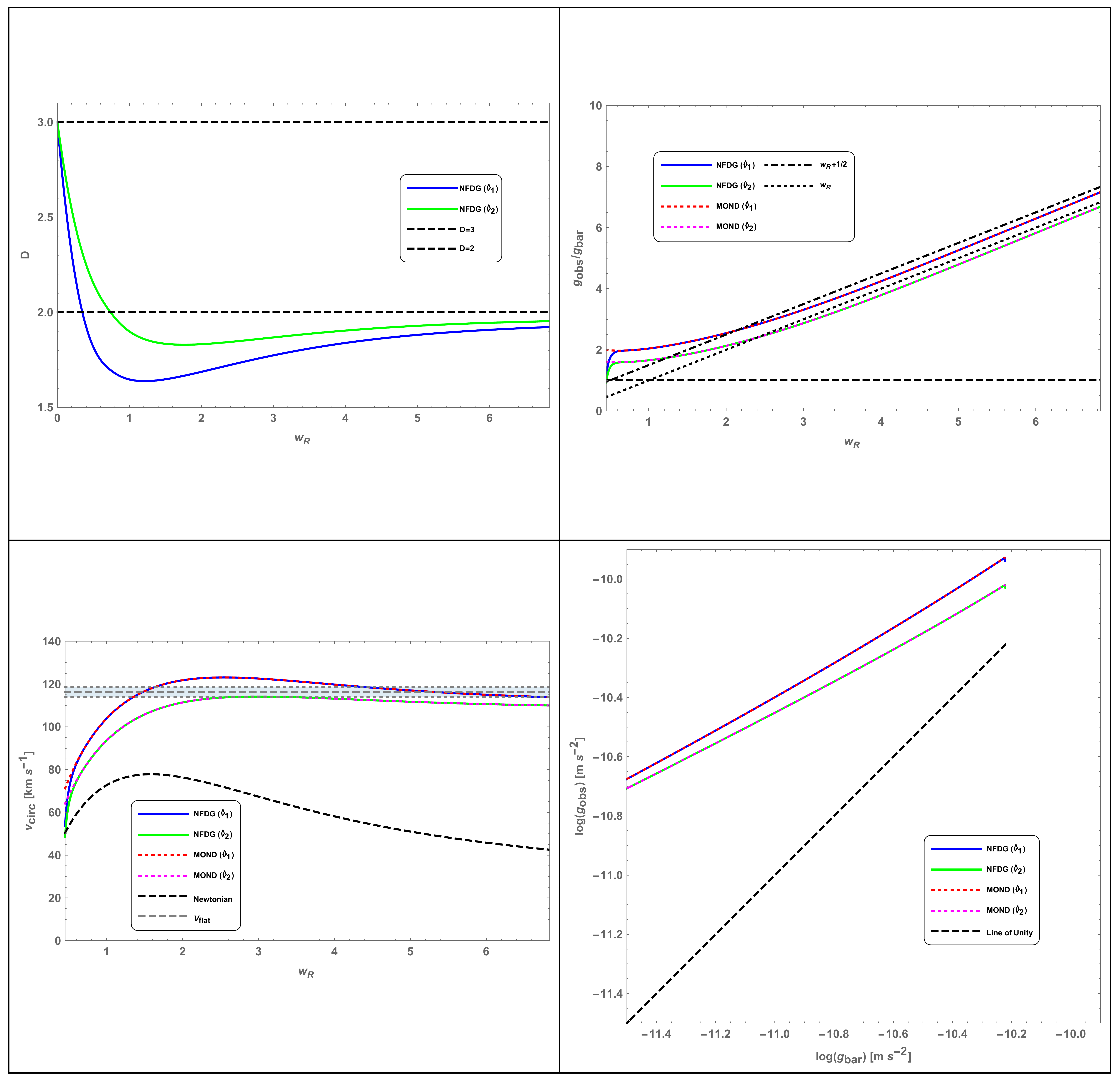}
}
\caption{Thick-disk exponential galaxy results.
Top-left panel: NFDG variable dimension $D\left (w_{R}\right )$ for MOND interpolation functions $\widehat{\nu }_{1}$ and $\widehat{\nu }_{2}$. Other panels: comparison of NFDG\ results (solid lines) with equivalent MOND predictions (dotted lines) for the two different interpolation functions. Also shown: Newtonian behavior-Line of Unity (black-dashed lines).}\label{figure:thickdiskexponential}\end{figure}

In our third case, we consider a thick-disk galaxy following the discussion in Sect. \ref{sect:AXIAL}. We use here the procedure based on Eqs. (\ref{eq3.18})-(\ref{eq3.19}) and described in the paragraphs after these two equations. The results shown in Fig. \ref{figure:thickdiskexponential} were obtained by performing a triple numerical integration and a summation over the first few non-zero terms of the kernel expansion, with the same physical parameters of NGC 6503 used before.

Due to the numerical integration procedure being used, the computation of $g_{obs}$ has some convergence issues at low $w_{R}$ values; therefore, we only show show results for $w_{R} \gtrsim 0.45$ in most panels of Fig. \ref{figure:thickdiskexponential}.\protect\footnote{
In the top-left panel of Fig. \ref{figure:thickdiskexponential}, we extrapolate the dimension functions below $w_{R} \approx 0.45$ by assuming $D =3$ at the origin.
} However, for the range of $w_{R}$ values shown in this figure, our model produces results that are very similar to those of the thin-disk case analyzed previously: we can only notice slight differences between the corresponding panels, comparing Fig. \ref{figure:thickdiskexponential} with Fig. \ref{figure:thindiskexponential}.

Therefore, we conclude that both thin/thick disk procedures can be used to analyze effectively disk galaxies:\ the thin-disk method is more efficient due to the fully analytical treatment, while the thick-disk procedure is more cumbersome, due to the numerical computations. In the section \ref{sect:6503}, we will use these methods for a complete fitting of the rotation curves of NGC 6503.

\subsection{\label{subsect:spherical}Spherical bulges
}
Although this paper is mostly devoted to the analysis of axially-symmetric structures, at the end of Sect. \ref{sect:AXIAL} we discussed how our approach, using a NFDG gravitational potential and the general expansion in Eq. (\ref{eq3.5}), can be adapted also to spherically-symmetric structures.

In paper I, we already studied the case of spherical symmetry in NFDG, but in that work our approach was based on Eq. (\ref{eq2.11}), i.e., a direct determination of the gravitational field $\mathbf{g}_{obs}\left (w_{r}\right )$ without computing the related potential $\phi \left (w_{r}\right )$. In this section, we will check that these two NFDG\ approaches to spherical structures are indeed equivalent, thus confirming that our methods are mathematically sound.

We will reconsider here one of the examples presented in our paper I: a simple Plummer model for a spherically symmetric structure as discussed in Sect. 4.4 of paper I. We recall that a Plummer gravitational potential is related to a rescaled mass density \cite{Varieschi:2020ioh}:

\begin{equation}\widetilde{\rho }\left (w_{r}^{ \prime }\right ) =\frac{3M}{4\pi W^{3}}\left (1 +\frac{w_{r}^{ \prime 2}}{W^{2}}\right )^{ -5/2} , \label{eq4.2}
\end{equation}where $M$ is the total mass and $W =b/l_{0}$ is the rescaled length of the Plummer original potential $\phi \left (r\right ) = -GM/\sqrt{r^{2} +b^{2}}$ \cite{2008gady.book.....B}. We will also use the same choice for the physical parameters as in paper I: $M \approx 2 \times 10^{5}M_{ \odot } \approx 4 \times 10^{35}\ \mbox{kg}$ and $W \approx 0.1$, which followed from typical data of globular clusters in our Galaxy (see again paper I for details).

\begin{figure}\centering 
\setlength\fboxrule{0in}\setlength\fboxsep{0.1in}\fcolorbox[HTML]{FFFFFF}{FFFFFF}{\includegraphics[ width=7.03in, height=6.7875862068965525in,]{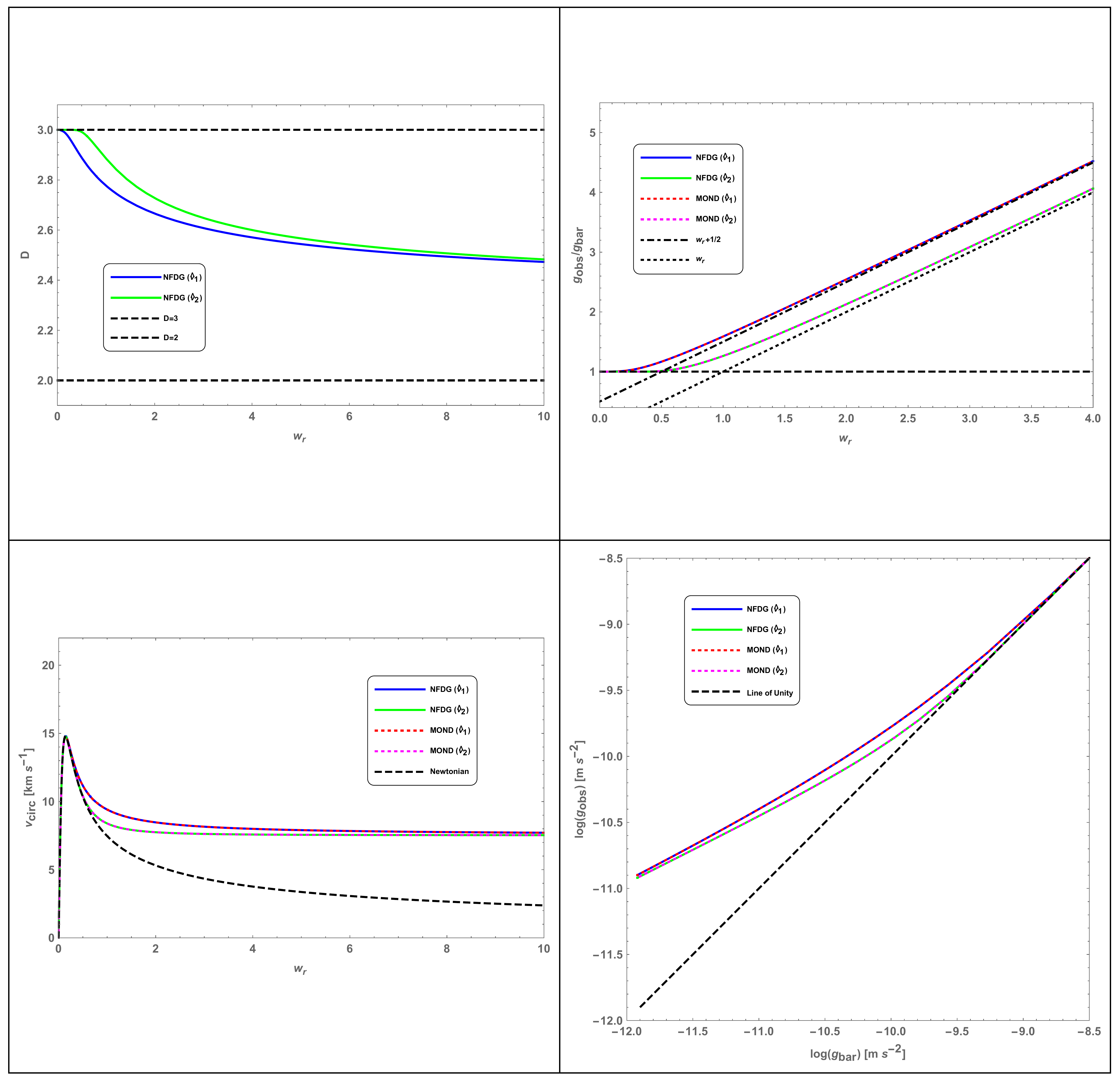}
}
\caption{Spherical Plummer model results.
Top-left panel: NFDG variable dimension $D\left (w_{r}\right )$ for MOND interpolation functions $\widehat{\nu }_{1}$ and $\widehat{\nu }_{2}$. Other panels: comparison of NFDG\ results (solid lines) with equivalent MOND predictions (dotted lines) for the two different interpolation functions. Also shown: Newtonian behavior-Line of Unity (black-dashed lines).
}\label{figure: Plummer}\end{figure}

 Figure \ref{figure: Plummer},
modeled after the first three figures, summarizes all results for this case. Once again, the top-left
panel shows dimension functions
$D\left (w_{r}\right )$
consistent with our NFDG model based on fractional gravity: the effective dimension decreases from the standard Newtonian
$D \approx 3$
value toward the deep-MOND
$D \approx 2$
value asymptotically.\protect\footnote{ In this figure, we limited the range of
$w_{r}$
between
$0$
and
$10$. Plotting the top-left panel for
$w_{r} \gg 10$ would show that
$D \rightarrow 2$
for large values of
$w_{r}$.
} This is also reflected in the top-right panel, by using the
$\genfrac{(}{)}{}{}{g_{obs}}{g_{bar}}$
ratios instead.\protect\footnote{
In this case, we set $\genfrac{(}{)}{}{}{g_{obs}}{g_{bar}} \rightarrow 1$ for $w_{r} \rightarrow 0$ (Newtonian behavior at the origin) as was done in paper I. Thus the ratio $\genfrac{(}{)}{}{}{g_{obs}}{g_{bar}}$ does not diverge for $w_{r} \rightarrow 0$, as in the previous cases analyzed in this paper.
} The two bottom panels in Fig. \ref{figure: Plummer}, also confirm that our NFDG model can yield the same results of the standard
MOND theory, but with circular speed plots and log-log plots now explained by our variable-dimension effect, as
opposed to just an empirical MOND-RAR relation.

Comparing this Fig. \ref{figure: Plummer} with figure 3 in our previous paper I, we can immediately see that the results are practically identical. However, figure 3 in paper I was obtained by using directly the NFDG\ field from Eq. (\ref{eq2.11}), while our current Fig. \ref{figure: Plummer} was obtained using the NFDG potential model, combining together Eqs. (\ref{eq3.23}), (\ref{eq3.24}), and (\ref{eq4.2}). This confirms that the NFDG methods developed in both paper I and the current work are mathematically consistent. In future studies, we will be able to use the methods described in this section for more general spherical structures, such as galactic spherical bulges, dwarf spheroidal galaxies, or others.

\section{\label{sect:6503}Spiral galaxy NGC 6503
}
Although we already used the astrophysical data of the field dwarf spiral galaxy NGC 6503 in Sect. \ref{sect::galactic}, in this section we want to show how NFDG\ methods can be applied directly to fitting rotation curves of a particular galaxy. The choice of NGC 6503 simply follows from the fact that this galaxy was used as one of the three main examples of the RAR in the seminal paper by McGaugh et al. \cite{McGaugh:2016leg}.

\subsection{\label{subsect:data}NGC 6503 data fitting
}
In the following, we will consider the full astrophysical data for NGC 6503 as available in the SPARC database \cite{Lelli:2016zqa,Lelli:2020pri}, and not just the two main parameters: mass $M =1.72 \times 10^{40}\thinspace \mbox{kg}$ and disk scale length $R_{d} =2.16\ \ensuremath{\operatorname*{kpc}} =6.67 \times 10^{19}\ \mbox{m}$, used in the previous sections as our main reference data. As the main input, we will use the disk and gas surface mass distributions, $\Sigma _{disk}\left (R\right )$  and $\Sigma _{gas}\left (R\right )$, respectively, which can be obtained from the SPARC surface luminosities $\Sigma _{disk}^{(L)}\left (R\right )$ and $\Sigma _{gas}^{(L)}\left (R\right )$, by using appropriate mass-to-light ratios \cite{Lelli:2020pri}: $\Upsilon _{disk} \simeq 0.50\ M_{ \odot }/L_{ \odot }$, $\Upsilon _{gas} \simeq 1.33\ M_{ \odot }/L_{ \odot }$ (this value for $\Upsilon _{gas}$ includes also the helium gas contribution). 

No spherical bulge is present for NGC 6503, so we will treat this galaxy as a thick-disk structure of total surface mass density $\Sigma \left (R\right ) =\Sigma _{disk}\left (R\right ) +\Sigma _{gas}\left (R\right )$ (obtained by interpolating the SPARC data), plus a vertical exponential density function $\widetilde{\zeta }_{1}$ from Eq. (\ref{eq3.19}), supplemented with the standard relation \cite{Lelli:2016zqa,2010ApJ...716..234B}, $\left (h_{z}/\ensuremath{\operatorname*{kpc}}\right ) =0.196\left (R_{d}/\ensuremath{\operatorname*{kpc}}\right )^{0.633}$, between the vertical/radial exponential scale lengths.

After rescaling all variables as usual, our NFDG\ ratio
$\genfrac{(}{)}{}{}{g_{obs}}{g_{bar}}_{NFDG}\left (w_{R}\right ) =\frac{\left \vert \mathbf{g}_{obs}\left (w_{R} ,D\left (w_{R}\right )\right )\right \vert }{\left \vert \mathbf{g}_{bar}\left (w_{R}\right )\right \vert }$ can be compared with the MOND-RAR ratio
$\genfrac{(}{)}{}{}{g_{obs}}{g_{bar}}_{MOND}\left (w_{R}\right ) =\widehat{\nu }_{1}\left (w_{R}\right ) =\frac{1}{1 -e^{ -\sqrt{g_{bar}\left (w_{R}\right )/g_{\dag }}}}$, or directly with the ratio of the experimental\ SPARC data for NGC 6503: $\genfrac{(}{)}{}{}{g_{obs}}{g_{bar}}_{SPARC}\left (w_{R}\right )$. This ratio is easily obtained from the published \cite{Lelli:2016zqa,Lelli:2020pri} rotation velocity data for NGC 6503 and allows for a more direct validation of our methods.

\begin{figure}\centering 
\setlength\fboxrule{0in}\setlength\fboxsep{0.1in}\fcolorbox[HTML]{FFFFFF}{FFFFFF}{\includegraphics[ width=7.05in, height=6.806896551724138in,]{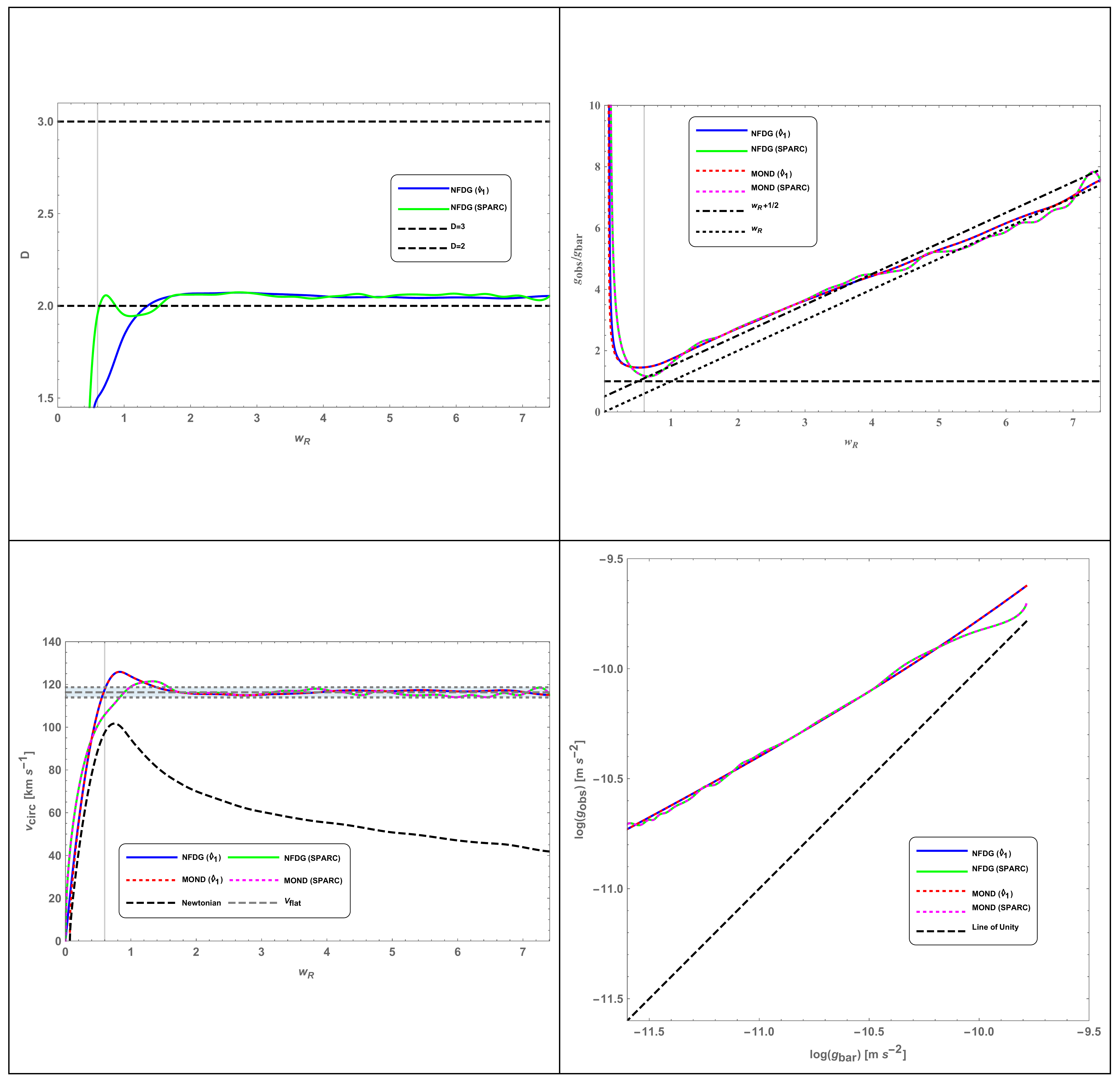}
}
\caption{NGC 6503 dwarf spiral galaxy results.
Top-left panel: NFDG variable dimension $D\left (w_{R}\right )$ for MOND interpolation function $\widehat{\nu }_{1}$ and using SPARC data. Other panels: comparison of NFDG\ results (solid lines) with equivalent MOND predictions (dotted lines), for the two different cases being considered. Also shown: Newtonian behavior-Line of Unity (black-dashed lines) and thin-gray vertical lines at $w_{R} \simeq 0.60$, indicating radial distance below which our results have been extrapolated.
}\label{figure:NGC6503fig1}\end{figure}

In Fig. \ref{figure:NGC6503fig1} we summarize our results, following the same format used in the previous figures. In the top-left panel, we show the variable dimension $D\left (w_{R}\right )$ obtained by setting $\genfrac{(}{)}{}{}{g_{obs}}{g_{bar}}_{NFDG}\left (w_{R}\right ) =\widehat{\nu }_{1}\left (w_{R}\right )$ and $\genfrac{(}{)}{}{}{g_{obs}}{g_{bar}}_{NFDG}\left (w_{R}\right ) =\genfrac{(}{)}{}{}{g_{obs}}{g_{bar}}_{SPARC}\left (w_{R}\right )$, respectively. This time we did not assume $D =3$ at the origin, and we note that our procedure converges well only for $w_{R} \gtrsim 0.60$, while it has trouble converging at lower values as already described in Sect. \ref{subsect:exponentialthick}, due to the triple numerical integration procedure. Therefore, in the first three panels of the figure, results for $w_{R} \lesssim 0.60$ were simply extrapolated and this particular value of $w_{R} \simeq 0.60$ is shown with vertical thin-gray lines in these panels. The upper limiting value of $w_{R} \simeq 7.41$ follows instead from the largest radial distance in the SPARC data for this galaxy.

In the top-left panel, we see that over the range $w_{R} \gtrsim 1$, where our numerical routines converge rapidly, the  dimension $D\left (w_{R}\right )$ in both cases remains very close to the value $D \simeq 2$. This is different from all the cases analyzed in Sect. \ref{sect::galactic}, where we were not modeling real astrophysical data, but we were just using standard mass distributions. This might be an indication that, for a thin/thick galaxy with no spherical bulge, the system of stars behaves as a fractal medium with an almost constant fractional dimension value of $D \simeq 2$. This is also in line with our original heuristic analysis presented in paper I, showing that the fundamental MOND results are recovered in NFDG by simply assuming $D \simeq 2$. 

As usual, the top-right panel considers the
$\genfrac{(}{)}{}{}{g_{obs}}{g_{bar}}$
ratios and shows that NFDG\ methods can reproduce MOND results consistently, except for the lowest values of $w_{R}$ where our numerical procedures become unreliable. In this panel, as well as in the following ones, it should be noted that the comparison between NFDG and MOND results using SPARC data is more relevant than the similar comparison with the interpolation function $\widehat{\nu }_{1}$. In the former case we compare directly our results with the astrophysical data of NGC 6503, while in the latter we use the RAR, which is an approximated interpolation between the asymptotic MOND behaviors, based on all possible SPARC data. Thus, in all panels of this figure the NFDG (SPARC) curves (green-solid), should be considered the direct fit to the galactic data being analyzed.

The bottom-left panel shows circular velocity plots corresponding to the
previously analyzed cases, and compared with the purely Newtonian case. In this panel, as well as in the
bottom-right one, we used the NGC 6503 total mass $M =1.72 \times 10^{40}\thinspace \mbox{kg}$, obtained by integrating the total SPARC mass distribution (disk plus gas),           with
$l_{0} \approx \sqrt{\frac{GM}{a_{0}}} \simeq 9.79 \times 10^{19}\mbox{m}$, and disk scale length $R_{d} =2.16\ \ensuremath{\operatorname*{kpc}} =6.67 \times 10^{19}\ \mbox{m}$ (rescaled length $W_{d} =R_{d}/l_{0} =0.681$). The NFDG circular speeds are
computed as
$v_{circ} =\sqrt{\left (g_{obs}\right )_{NFDG}\left (w_{R} ,D\left (w_{R}\right )\right )w_{R}\thinspace l_{0}}/10^{3}$$\left [\mbox{km}\thinspace \mbox{s}^{ -1}\right ]$,
with the dimension functions $D\left (w_{R}\right )$ as obtained in the top-left panel. The MOND circular speeds are computed as
$v_{circ} =\sqrt{\left (g_{bar}\right )_{SPARC}\left (w_{R}\right )\widehat{\nu }_{1}\left (w_{R}\right )w_{R}\thinspace l_{0}}/10^{3}\left [\mbox{km}\thinspace \mbox{s}^{ -1}\right ]$
and $v_{circ} =\sqrt{\left (g_{obs}\right )_{SPARC}\left (w_{R}\right )w_{R}\thinspace l_{0}}/10^{3}\left [\mbox{km}\thinspace \mbox{s}^{ -1}\right ]$, respectively, while the purely Newtonian speed is
$v_{circ} =\sqrt{\left (g_{bar}\right )_{SPARC}\left (w_{R}\right )w_{R}\thinspace l_{0}}/10^{3}\left [\mbox{km}\thinspace \mbox{s}^{ -1}\right ]$.

As seen in this panel, there is perfect agreement between the respective - $\widehat{\nu }_{1}$
or SPARC - NFDG and MOND cases, showing the expected flattening of the circular speed plots over most of the $w_{R}$ range, as opposed to the standard Newtonian decrease of circular speed with radial distance. Again, the more direct fit to NGC 6503 data is represented by the NGC (SPARC) curve in green-solid. We also show the NGC 6503 flat rotation velocity
$V_{f} =116.3 \pm 2.4\left [\mbox{km}\thinspace \mbox{s}^{ -1}\right ]$  \cite{Lelli:2016zqa}, represented by the horizontal gray lines and band, as in the previous figures. Both NFDG (SPARC) and NFDG $\left (\widehat{\nu }_{1}\right )$ curves are well in agreement with these $V_{f}$ values for $w_{R} \gtrsim 1.5$, thus showing that NFDG methods can fully explain the main feature of the rotation velocity curves.

The final panel (bottom-right) summarizes as usual the results in terms of the customary log-log plots, compared with the line of unity, illustrating again that the non-linearities of the MOND model can be well explained by our approach based on a linear model with a variable local effective dimension. All the results in Fig. \ref{figure:NGC6503fig1} were plotted as functions of the rescaled radial coordinate $w_{R}$, as was done in the previous examples. In the following section \ref{subsect:discussion}, we will plot again the rotation velocity curve for NGC 6503 in terms of the standard radial coordinate $R$ together with the complete SPARC data for this galaxy, and we will attempt to outline some general considerations about the possible meaning of the variable dimension $D$.

\subsection{\label{subsect:discussion}Discussion}
\begin{figure}\centering 
\setlength\fboxrule{0in}\setlength\fboxsep{0.1in}\fcolorbox[HTML]{FFFFFF}{FFFFFF}{\includegraphics[ width=6.98in, height=4.425255220417634in,]{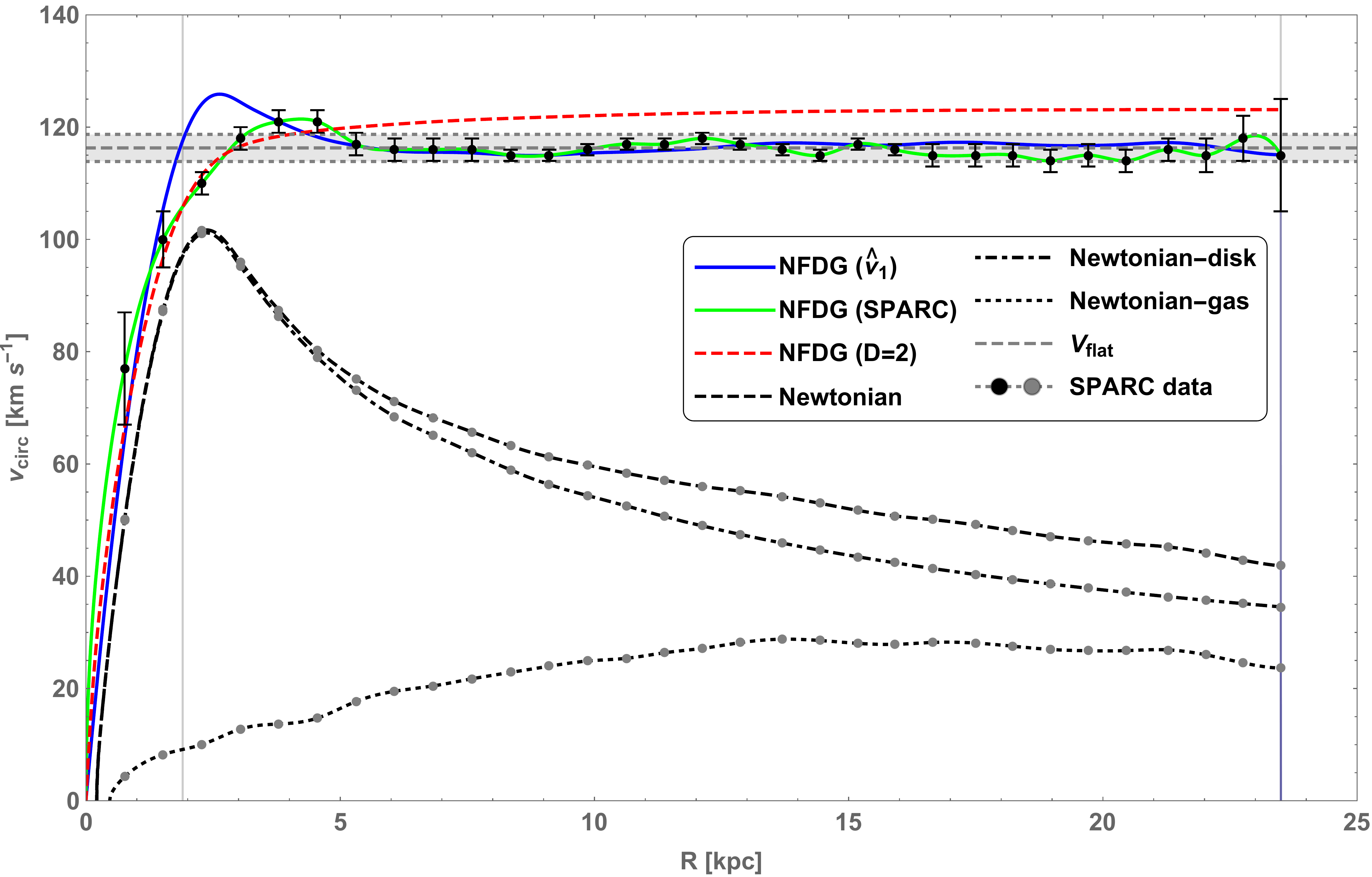}
}
\caption{NGC 6503 disk-dominated spiral galaxy rotation curves. NFDG fitting curves ($\widehat{\nu }_{1}$, SPARC, $D =2$) are compared directly with SPARC data-points (black circles) and with the flat rotation velocity $V_{f}$ (gray band). Also shown: Newtonian (total, disk, gas) rotation curves compared with original SPARC\ data (gray circles). Thin-gray vertical lines indicate the radial limits of our main NFDG fits.
}\label{figure:NGC6503fig2}\end{figure}

In figure \ref{figure:NGC6503fig2}, we produce our most detailed fitting to the NGC 6503 rotation velocity data, by expanding results already plotted in the bottom-left panel of Fig. \ref{figure:NGC6503fig1} and by adding the related SPARC data-points. We use here the physical radial distance $R$ in kiloparsec, while rotation circular velocities $v_{circ}$ are measured in $\mbox{km}\ \mbox{s}^{ -1}$.

In particular, in this figure we show the same NFDG (SPARC) and NFDG $\left (\widehat{\nu }_{1}\right )$ curves (green-solid and blue-solid, respectively) as in the bottom-left panel of the previous Fig. \ref{figure:NGC6503fig1}, which have been extrapolated below $R_{\min } \simeq 1.90\ \ kpc$, due to the convergence issues already mentioned. These curves are also limited by $R_{\max } \simeq 23.5\ kpc$, which is the radial distance of the last SPARC data-point. These limits are shown as vertical thin-gray lines in the figure. In addition, a third NFDG curve (red-dashed) is shown for a fixed value ($D =2$) of the space dimension. This curve can be computed even at very low radial distances, because it is based on the logarithmic potential in the second line of Eq. (\ref{eq2.3}), so it does not suffer from the numerical limitations at low-$R$ of the other NFDG curves.

These three NFDG curves can be compared directly with the SPARC data-points and related error bars (black circles), obtained from the published data \cite{Lelli:2016zqa,Lelli:2020pri}, and also with the flat rotation velocity
$V_{f} =116.3 \pm 2.4\left [\mbox{km}\thinspace \mbox{s}^{ -1}\right ]$  \cite{Lelli:2016zqa}, represented by the horizontal gray lines/band. For completeness, we also show the SPARC\  data for the Newtonian cases (disk contribution, gas contribution, and total Newtonian - gray circles) together with the computed Newtonian curves (in gray) from the extrapolated mass distributions (derived from the original luminosity distributions \cite{Lelli:2020pri}).

For this disk-dominated spiral galaxy, the flattening effect of the observed rotation curve is evident over most of the radial range. Our NFDG (SPARC) curve (green-solid line) can perfectly model the published data over the applicable range ($R_{\min } ,R_{\max }$) and even at the lower radial values, where the NFDG\ results have been extrapolated. Again, this green-solid curve is obtained by assuming the variable dimension $D\left (w_{R}\right )$ as described by the corresponding green-solid curve in the top-left panel of Fig. \ref{figure:NGC6503fig1}, i.e., assuming that NGC 6503 behaves as a fractal medium whose fractional dimension is described by this function $D\left (w_{R}\right )$.

The  NFDG $\left (\widehat{\nu }_{1}\right )$ curve (blue-solid line) is less effective than the previous one in modeling the SPARC data, but still well within the gray band of the flat rotation velocities over most of the radial range. As already remarked, this curve corresponds to the general RAR, which is an empirical fit to all SPARC\  data-points and, therefore, less accurate in the analysis of an individual galaxy. We also plot the third NFDG curve (red-dashed) for a fixed value ($D =2$) because this can be easily computed even at low values of $R$. This $D =2$ curve is less effective over most of the radial range (although remarkably flat), but it shows a good fit to the first few SPARC data-points at low radial distances, where our general methods are not completely reliable.

We interpret the above results as a possible indication that, for a disk-dominated galaxy like NGC 6503 or similar, the fractal dimension over the whole radial range is approximately $D \simeq 2$, with just small variations from this almost constant value. This is in line with all our heuristic arguments, presented in this paper as well as in paper I, that the MOND\ behavior is essentially related to an effective Newtonian gravity in a space of reduced dimension $D \simeq 2$. This could simply be the consequence of the shape, or other geometrical characteristics, of highly flattened structures such as thin/thich disk galaxies, as opposed to more spherical structures, such as globular clusters, galaxy clusters, etc., where MOND\ is much less effective.

It should be noted that the results for the varying dimension $D$ of NGC 6503 shown in the top-left panel of Fig. \ref{figure:NGC6503fig1}, as well as those previously illustrated in the similar top-left panels of Figs. \ref{figure:thindiskexponential}-\ref{figure: Plummer}, should be interpreted as the dimension of the matter distribution of the galactic structure being considered and not as the local space dimension that an observer would measure at a specific galactic location. In other words, if applied to our Galaxy the NFDG model would not contradict the tight bounds on the space dimension set by the Cassini measurements in the solar system \cite{Will:2014kxa}.

Further work will be needed to check the interpretation of all the results presented in this paper. This future work will include detailed fitting of structures where the MOND model is highly efffective (spiral, elliptical, and irregular galaxies), possibly showing a similar $2 \lesssim D \lesssim 3$ reduced dimension for these cases, as well as structures where MOND is not very effective (globular clusters, and similar spherical structures). In these latter cases, NFDG might still be able to describe the astrophysical data without using DM, but considering instead minor changes of the space dimension around $D \simeq 3$, which might account for the observed non-Newtonian behavior.

\section{\label{sect::conclusion}
Conclusion
}
In this work, we continued our study of a possible explanation of the MOND theory and related RAR
in terms of a novel fractional-dimension gravity model. As in our paper I, we considered the possibility that Newtonian gravity
might act on a metric space of variable dimension
$D \leq 3$, when applied to galactic scales, and developed the mathematical bases of NFDG for axially-symmetric galaxies.

The MOND acceleration scale
$a_{0}$, or the equivalent RAR acceleration parameter
$g_{\dag } =1.20 \times 10^{ -10}\mbox{m}\thinspace \mbox{s}^{ -2}$, were related to a length scale
$l_{0} \approx \sqrt{\frac{GM}{a_{0}}}$ which is naturally required for dimensional reasons and the NFDG gravitational potential was used as the key element of our model, together with the appropriate expansion in Gegenbauer polynomials of the Euler-NFDG kernel.

Using these assumptions, we have shown that, even in the case of axially-symmetric structures, our
NFDG can reproduce the same results of the MOND-RAR models and that the deep-MOND limit can be
achieved by continuously decreasing the space dimension
$D$
toward a limiting value of
$D \approx 2$. These methods were successfully applied to several different general cases (thin/thick exponential disks, Kuzmin disks, spherical bulges), and also to the detailed fitting of rotation data for the spiral disk galaxy NGC 6503. We have also considered a possible origin of the continuous
variation of the space dimension
$D$
for the case of NGC 6503, simply noting that over the whole deep-MOND\ regime the dimensional value might naturally approach $D \approx 2$, as this might be a common feature of all thin/thick disk galaxies.

Future work on the subject will
still be needed to test all these NFDG hypotheses. At this point, we have developed most of the mathematical tools to be used for detailed fitting of galactic rotation curves for any type of structure (thin/thick disks, spherical bulges, etc.). More detailed galactic fits will need to be performed, for several other galaxies in the SPARC database, before NFDG
can be considered a viable alternative model. In particular, more individual disk galaxies need to be analyzed in order to confirm our assumption that a natural value of $D \approx 2$ applies to these galaxies. Conversely, other structures such as globular clusters or similar need to be studied, including objects for which MOND does not seem to fully apply, but whose dynamical behavior might still be explained by NFDG without any use of DM.

 Lastly, a relativistic version of Newtonian Fractional-Dimension Gravity also needs to be established, possibly leading to an extension of General Relativity to metric spaces with fractional dimension. This relativistic version might also explain the possible origin of the variable dimension for the galactic structures studied in this paper. So far, we have been unable to derive NFDG from first principles and we have used it as an ad hoc model to fit the existing phenomenology. A fully relativistic version might be able to expand NFDG into a more fundamental theory. We will leave these and other topics to future work on the subject.

\begin{acknowledgments}This work was supported by a Faculty Sabbatical Leave granted by Loyola Marymount University, Los Angeles. The author also wishes to
acknowledge Dr. Howard Cohl for his advice regarding computations with
special functions, Dr. Federico Lelli for sharing NGC 6503 galactic data files and other useful information, and the anonymous reviewer for useful comments and suggestions.
\end{acknowledgments}
\bibliographystyle{apsrev4-1}
\bibliography{mainNotes}

\begin{thebibliography}{63}%
\makeatletter
\providecommand \@ifxundefined [1]{%
 \@ifx{#1\undefined}
}%
\providecommand \@ifnum [1]{%
 \ifnum #1\expandafter \@firstoftwo
 \else \expandafter \@secondoftwo
 \fi
}%
\providecommand \@ifx [1]{%
 \ifx #1\expandafter \@firstoftwo
 \else \expandafter \@secondoftwo
 \fi
}%
\providecommand \natexlab [1]{#1}%
\providecommand \enquote  [1]{``#1''}%
\providecommand \bibnamefont  [1]{#1}%
\providecommand \bibfnamefont [1]{#1}%
\providecommand \citenamefont [1]{#1}%
\providecommand \href@noop [0]{\@secondoftwo}%
\providecommand \href [0]{\begingroup \@sanitize@url \@href}%
\providecommand \@href[1]{\@@startlink{#1}\@@href}%
\providecommand \@@href[1]{\endgroup#1\@@endlink}%
\providecommand \@sanitize@url [0]{\catcode `\\12\catcode `\$12\catcode
  `\&12\catcode `\#12\catcode `\^12\catcode `\_12\catcode `\%12\relax}%
\providecommand \@@startlink[1]{}%
\providecommand \@@endlink[0]{}%
\providecommand \url  [0]{\begingroup\@sanitize@url \@url }%
\providecommand \@url [1]{\endgroup\@href {#1}{\urlprefix }}%
\providecommand \urlprefix  [0]{URL }%
\providecommand \Eprint [0]{\href }%
\providecommand \doibase [0]{http://dx.doi.org/}%
\providecommand \selectlanguage [0]{\@gobble}%
\providecommand \bibinfo  [0]{\@secondoftwo}%
\providecommand \bibfield  [0]{\@secondoftwo}%
\providecommand \translation [1]{[#1]}%
\providecommand \BibitemOpen [0]{}%
\providecommand \bibitemStop [0]{}%
\providecommand \bibitemNoStop [0]{.\EOS\space}%
\providecommand \EOS [0]{\spacefactor3000\relax}%
\providecommand \BibitemShut  [1]{\csname bibitem#1\endcsname}%
\let\auto@bib@innerbib\@empty
%</preamble>
\bibitem [{\citenamefont {Varieschi}(2020{\natexlab{a}})}]{Varieschi:2020ioh}%
  \BibitemOpen
  \bibfield  {author} {\bibinfo {author} {\bibfnamefont {G.~U.}\ \bibnamefont
  {Varieschi}},\ }\href {\doibase 10.1007/s10701-020-00389-7} {\bibfield
  {journal} {\bibinfo  {journal} {Found. Phys.}\ }\textbf {\bibinfo {volume}
  {50}},\ \bibinfo {pages} {1608} (\bibinfo {year} {2020}{\natexlab{a}})},\
  \Eprint {http://arxiv.org/abs/2003.05784} {arXiv:2003.05784 [gr-qc]}
  \BibitemShut {NoStop}%
\bibitem [{\citenamefont {Varieschi}(2020{\natexlab{b}})}]{Varieschi:2020}%
  \BibitemOpen
  \bibfield  {author} {\bibinfo {author} {\bibfnamefont {G.~U.}\ \bibnamefont
  {Varieschi}},\ }\href {http://gvarieschi.lmu.build/NFDG2020.html} {\emph
  {\bibinfo {title} {Newtonian Fractional-Dimension Gravity (NFDG),
  http://gvarieschi.lmu.build/NFDG2020.html}}} (\bibinfo {year}
  {2020}{\natexlab{b}})\BibitemShut {NoStop}%
\bibitem [{\citenamefont {Tarasov}(2011)}]{bookTarasov}%
  \BibitemOpen
  \bibfield  {author} {\bibinfo {author} {\bibfnamefont {V.}~\bibnamefont
  {Tarasov}},\ }\href@noop {} {\emph {\bibinfo {title} {Fractional Dynamics:
  Application of Fractional Calculus to Dynamics of Particles, Fields and
  Media}}}\ (\bibinfo {year} {2011})\BibitemShut {NoStop}%
\bibitem [{\citenamefont {Zubair}\ \emph {et~al.}(2012)\citenamefont {Zubair},
  \citenamefont {Mughal},\ and\ \citenamefont {Naqvi}}]{bookZubair}%
  \BibitemOpen
  \bibfield  {author} {\bibinfo {author} {\bibfnamefont {M.}~\bibnamefont
  {Zubair}}, \bibinfo {author} {\bibfnamefont {M.}~\bibnamefont {Mughal}}, \
  and\ \bibinfo {author} {\bibfnamefont {Q.}~\bibnamefont {Naqvi}},\ }\href
  {\doibase 10.1007/978-3-642-25358-4} {\emph {\bibinfo {title}
  {Electromagnetic Fields and Waves in Fractional Dimensional Space}}}\
  (\bibinfo {year} {2012})\BibitemShut {NoStop}%
\bibitem [{\citenamefont {Varieschi}(2018)}]{Varieschi:2018}%
  \BibitemOpen
  \bibfield  {author} {\bibinfo {author} {\bibfnamefont {G.~U.}\ \bibnamefont
  {Varieschi}},\ }\href {\doibase 10.4236/jamp.2018.66105} {\bibfield
  {journal} {\bibinfo  {journal} {J. Appl. Math.Phys.}\ }\textbf {\bibinfo
  {volume} {06}},\ \bibinfo {pages} {1247} (\bibinfo {year} {2018})},\ \Eprint
  {http://arxiv.org/abs/1712.03473} {arXiv:1712.03473 [physics.class-ph]}
  \BibitemShut {NoStop}%
\bibitem [{\citenamefont {Oldham}\ and\ \citenamefont
  {Spanier}(1974)}]{MR0361633}%
  \BibitemOpen
  \bibfield  {author} {\bibinfo {author} {\bibfnamefont {K.~B.}\ \bibnamefont
  {Oldham}}\ and\ \bibinfo {author} {\bibfnamefont {J.}~\bibnamefont
  {Spanier}},\ }\href@noop {} {\emph {\bibinfo {title} {The fractional
  calculus}}}\ (\bibinfo  {publisher} {Academic Press [A subsidiary of Harcourt
  Brace Jovanovich, Publishers], New York-London},\ \bibinfo {year} {1974})\
  pp.\ \bibinfo {pages} {xiii+234},\ \bibinfo {note} {theory and applications
  of differentiation and integration to arbitrary order, with an annotated
  chronological bibliography by Bertram Ross, Mathematics in Science and
  Engineering, Vol. 111}\BibitemShut {NoStop}%
\bibitem [{\citenamefont {Miller}\ and\ \citenamefont
  {Ross}(1993)}]{MR1219954}%
  \BibitemOpen
  \bibfield  {author} {\bibinfo {author} {\bibfnamefont {K.~S.}\ \bibnamefont
  {Miller}}\ and\ \bibinfo {author} {\bibfnamefont {B.}~\bibnamefont {Ross}},\
  }\href@noop {} {\emph {\bibinfo {title} {An introduction to the fractional
  calculus and fractional differential equations}}},\ A Wiley-Interscience
  Publication\ (\bibinfo  {publisher} {John Wiley \& Sons, Inc., New York},\
  \bibinfo {year} {1993})\ pp.\ \bibinfo {pages} {xvi+366}\BibitemShut
  {NoStop}%
\bibitem [{\citenamefont {Podlubny}(1999)}]{MR1658022}%
  \BibitemOpen
  \bibfield  {author} {\bibinfo {author} {\bibfnamefont {I.}~\bibnamefont
  {Podlubny}},\ }\href@noop {} {\emph {\bibinfo {title} {Fractional
  differential equations}}},\ \bibinfo {series} {Mathematics in Science and
  Engineering}, Vol.\ \bibinfo {volume} {198}\ (\bibinfo  {publisher} {Academic
  Press, Inc., San Diego, CA},\ \bibinfo {year} {1999})\ pp.\ \bibinfo {pages}
  {xxiv+340},\ \bibinfo {note} {an introduction to fractional derivatives,
  fractional differential equations, to methods of their solution and some of
  their applications}\BibitemShut {NoStop}%
\bibitem [{\citenamefont {Herrmann}(2011)}]{Herrmann:2011zza}%
  \BibitemOpen
  \bibfield  {author} {\bibinfo {author} {\bibfnamefont {R.}~\bibnamefont
  {Herrmann}},\ }\href@noop {} {\emph {\bibinfo {title} {Fractional calculus:
  An introduction for physicists}}}\ (\bibinfo  {publisher} {World Scientific,
  Hackensack, USA},\ \bibinfo {year} {2011})\ p.\ \bibinfo {pages}
  {261}\BibitemShut {NoStop}%
\bibitem [{\citenamefont {Hilfer}(2000)}]{MR1890104}%
  \BibitemOpen
  \bibinfo {editor} {\bibfnamefont {R.}~\bibnamefont {Hilfer}},\ ed.,\ \href
  {\doibase 10.1142/9789812817747} {\emph {\bibinfo {title} {Applications of
  fractional calculus in physics}}}\ (\bibinfo  {publisher} {World Scientific
  Publishing Co., Inc., River Edge, NJ},\ \bibinfo {year} {2000})\ pp.\
  \bibinfo {pages} {viii+463}\BibitemShut {NoStop}%
\bibitem [{\citenamefont {Baryshev}\ and\ \citenamefont
  {Teerikorpi}(2002)}]{Baryshev:2002tn}%
  \BibitemOpen
  \bibfield  {author} {\bibinfo {author} {\bibfnamefont {{\relax
  Yu}.}~\bibnamefont {Baryshev}}\ and\ \bibinfo {author} {\bibfnamefont
  {P.}~\bibnamefont {Teerikorpi}},\ }\href@noop {} {\emph {\bibinfo {title}
  {{Discovery of cosmic fractals}}}}\ (\bibinfo  {publisher} {River Edge, USA:
  World Scientific (2002) 373 p},\ \bibinfo {year} {2002})\BibitemShut
  {NoStop}%
%%CITATION = INSPIRE-607412;%%
\bibitem [{\citenamefont {Nottale}(2011)}]{Nottale:2011zz}%
  \BibitemOpen
  \bibfield  {author} {\bibinfo {author} {\bibfnamefont {L.}~\bibnamefont
  {Nottale}},\ }\href {http://www.icpress.co.uk/physics/p752.html} {\emph
  {\bibinfo {title} {{Scale relativity and fractal space-time}}}}\ (\bibinfo
  {publisher} {London, UK: Imp. Coll. Pr. (2011) 742 p},\ \bibinfo {year}
  {2011})\BibitemShut {NoStop}%
%%CITATION = INSPIRE-1083945;%%
\bibitem [{\citenamefont {Calcagni}(2017{\natexlab{a}})}]{Calcagni:2016azd}%
  \BibitemOpen
  \bibfield  {author} {\bibinfo {author} {\bibfnamefont {G.}~\bibnamefont
  {Calcagni}},\ }\href {\doibase 10.1007/JHEP03(2017)138,
  10.1007/JHEP06(2017)020} {\bibfield  {journal} {\bibinfo  {journal} {JHEP}\
  }\textbf {\bibinfo {volume} {03}},\ \bibinfo {pages} {138} (\bibinfo {year}
  {2017}{\natexlab{a}})},\ \Eprint {http://arxiv.org/abs/1612.05632}
  {arXiv:1612.05632 [hep-th]} \BibitemShut {NoStop}%
%%CITATION = ARXIV:1612.05632;%%
\bibitem [{\citenamefont {{Muslih}}\ \emph {et~al.}(2007)\citenamefont
  {{Muslih}}, \citenamefont {{Baleanu}},\ and\ \citenamefont
  {{Rabei}}}]{Muslih2007}%
  \BibitemOpen
  \bibfield  {author} {\bibinfo {author} {\bibfnamefont {S.~I.}\ \bibnamefont
  {{Muslih}}}, \bibinfo {author} {\bibfnamefont {D.}~\bibnamefont {{Baleanu}}},
  \ and\ \bibinfo {author} {\bibfnamefont {E.~M.}\ \bibnamefont {{Rabei}}},\
  }\href {\doibase 10.2478/s11534-007-0014-9} {\bibfield  {journal} {\bibinfo
  {journal} {Central European Journal of Physics}\ }\textbf {\bibinfo {volume}
  {5}},\ \bibinfo {pages} {285} (\bibinfo {year} {2007})}\BibitemShut {NoStop}%
\bibitem [{\citenamefont {{Rousan}}\ \emph {et~al.}(2002)\citenamefont
  {{Rousan}}, \citenamefont {{Malkawi}}, \citenamefont {{Rabei}},\ and\
  \citenamefont {{Widyan}}}]{Rousan2002}%
  \BibitemOpen
  \bibfield  {author} {\bibinfo {author} {\bibfnamefont {A.~A.}\ \bibnamefont
  {{Rousan}}}, \bibinfo {author} {\bibfnamefont {E.}~\bibnamefont {{Malkawi}}},
  \bibinfo {author} {\bibfnamefont {E.~M.}\ \bibnamefont {{Rabei}}}, \ and\
  \bibinfo {author} {\bibfnamefont {H.}~\bibnamefont {{Widyan}}},\ }\href@noop
  {} {\bibfield  {journal} {\bibinfo  {journal} {Frac. Calc. Appl. Anal.}\
  }\textbf {\bibinfo {volume} {5}},\ \bibinfo {pages} {155} (\bibinfo {year}
  {2002})}\BibitemShut {NoStop}%
\bibitem [{\citenamefont {Munkhammar}(2010)}]{Munkhammar:2010gq}%
  \BibitemOpen
  \bibfield  {author} {\bibinfo {author} {\bibfnamefont {J.}~\bibnamefont
  {Munkhammar}},\ }\href@noop {} {\  (\bibinfo {year} {2010})},\ \Eprint
  {http://arxiv.org/abs/1003.4981} {arXiv:1003.4981 [physics.gen-ph]}
  \BibitemShut {NoStop}%
\bibitem [{\citenamefont {Calcagni}(2010)}]{Calcagni:2009kc}%
  \BibitemOpen
  \bibfield  {author} {\bibinfo {author} {\bibfnamefont {G.}~\bibnamefont
  {Calcagni}},\ }\href {\doibase 10.1103/PhysRevLett.104.251301} {\bibfield
  {journal} {\bibinfo  {journal} {Phys. Rev. Lett.}\ }\textbf {\bibinfo
  {volume} {104}},\ \bibinfo {pages} {251301} (\bibinfo {year} {2010})},\
  \Eprint {http://arxiv.org/abs/0912.3142} {arXiv:0912.3142 [hep-th]}
  \BibitemShut {NoStop}%
%%CITATION = ARXIV:0912.3142;%%
\bibitem [{\citenamefont {Calcagni}(2012{\natexlab{a}})}]{Calcagni:2011kn}%
  \BibitemOpen
  \bibfield  {author} {\bibinfo {author} {\bibfnamefont {G.}~\bibnamefont
  {Calcagni}},\ }\href {\doibase 10.4310/ATMP.2012.v16.n2.a5} {\bibfield
  {journal} {\bibinfo  {journal} {Adv. Theor. Math. Phys.}\ }\textbf {\bibinfo
  {volume} {16}},\ \bibinfo {pages} {549} (\bibinfo {year}
  {2012}{\natexlab{a}})},\ \Eprint {http://arxiv.org/abs/1106.5787}
  {arXiv:1106.5787 [hep-th]} \BibitemShut {NoStop}%
%%CITATION = ARXIV:1106.5787;%%
\bibitem [{\citenamefont {Calcagni}(2012{\natexlab{b}})}]{Calcagni:2011sz}%
  \BibitemOpen
  \bibfield  {author} {\bibinfo {author} {\bibfnamefont {G.}~\bibnamefont
  {Calcagni}},\ }\href {\doibase 10.1007/JHEP01(2012)065} {\bibfield  {journal}
  {\bibinfo  {journal} {JHEP}\ }\textbf {\bibinfo {volume} {01}},\ \bibinfo
  {pages} {065} (\bibinfo {year} {2012}{\natexlab{b}})},\ \Eprint
  {http://arxiv.org/abs/1107.5041} {arXiv:1107.5041 [hep-th]} \BibitemShut
  {NoStop}%
%%CITATION = ARXIV:1107.5041;%%
\bibitem [{\citenamefont {Calcagni}(2013)}]{Calcagni:2013yqa}%
  \BibitemOpen
  \bibfield  {author} {\bibinfo {author} {\bibfnamefont {G.}~\bibnamefont
  {Calcagni}},\ }\href {\doibase 10.1088/1475-7516/2013/12/041} {\bibfield
  {journal} {\bibinfo  {journal} {JCAP}\ }\textbf {\bibinfo {volume} {1312}},\
  \bibinfo {pages} {041} (\bibinfo {year} {2013})},\ \Eprint
  {http://arxiv.org/abs/1307.6382} {arXiv:1307.6382 [hep-th]} \BibitemShut
  {NoStop}%
%%CITATION = ARXIV:1307.6382;%%
\bibitem [{\citenamefont {Calcagni}(2017{\natexlab{b}})}]{Calcagni:2016xtk}%
  \BibitemOpen
  \bibfield  {author} {\bibinfo {author} {\bibfnamefont {G.}~\bibnamefont
  {Calcagni}},\ }\href {\doibase 10.1103/PhysRevD.95.064057} {\bibfield
  {journal} {\bibinfo  {journal} {Phys. Rev. D}\ }\textbf {\bibinfo {volume}
  {95}},\ \bibinfo {pages} {064057} (\bibinfo {year} {2017}{\natexlab{b}})},\
  \Eprint {http://arxiv.org/abs/1609.02776} {arXiv:1609.02776 [gr-qc]}
  \BibitemShut {NoStop}%
\bibitem [{\citenamefont {Calcagni}(2018)}]{Calcagni:2018dhp}%
  \BibitemOpen
  \bibfield  {author} {\bibinfo {author} {\bibfnamefont {G.}~\bibnamefont
  {Calcagni}},\ }\href {\doibase 10.3389/fphy.2018.00058} {\bibfield  {journal}
  {\bibinfo  {journal} {Front.in Phys.}\ }\textbf {\bibinfo {volume} {6}},\
  \bibinfo {pages} {58} (\bibinfo {year} {2018})},\ \Eprint
  {http://arxiv.org/abs/1801.00396} {arXiv:1801.00396 [math-ph]} \BibitemShut
  {NoStop}%
%%CITATION = ARXIV:1801.00396;%%
\bibitem [{\citenamefont {Svozil}(2017)}]{Svozil:2017ybx}%
  \BibitemOpen
  \bibfield  {author} {\bibinfo {author} {\bibfnamefont {K.}~\bibnamefont
  {Svozil}},\ }\href@noop {} {\  (\bibinfo {year} {2017})},\ \Eprint
  {http://arxiv.org/abs/1712.01376} {arXiv:1712.01376 [physics.class-ph]}
  \BibitemShut {NoStop}%
\bibitem [{\citenamefont {Giusti}(2020)}]{Giusti:2020rul}%
  \BibitemOpen
  \bibfield  {author} {\bibinfo {author} {\bibfnamefont {A.}~\bibnamefont
  {Giusti}},\ }\href {\doibase 10.1103/PhysRevD.101.124029} {\bibfield
  {journal} {\bibinfo  {journal} {Phys. Rev. D}\ }\textbf {\bibinfo {volume}
  {101}},\ \bibinfo {pages} {124029} (\bibinfo {year} {2020})},\ \Eprint
  {http://arxiv.org/abs/2002.07133} {arXiv:2002.07133 [gr-qc]} \BibitemShut
  {NoStop}%
\bibitem [{\citenamefont {Giusti}\ \emph {et~al.}(2020)\citenamefont {Giusti},
  \citenamefont {Garrappa},\ and\ \citenamefont {Vachon}}]{Giusti:2020kcv}%
  \BibitemOpen
  \bibfield  {author} {\bibinfo {author} {\bibfnamefont {A.}~\bibnamefont
  {Giusti}}, \bibinfo {author} {\bibfnamefont {R.}~\bibnamefont {Garrappa}}, \
  and\ \bibinfo {author} {\bibfnamefont {G.}~\bibnamefont {Vachon}},\ }\href
  {\doibase 10.1140/epjp/s13360-020-00831-9} {\bibfield  {journal} {\bibinfo
  {journal} {Eur. Phys. J. Plus}\ }\textbf {\bibinfo {volume} {135}},\ \bibinfo
  {pages} {798} (\bibinfo {year} {2020})},\ \Eprint
  {http://arxiv.org/abs/2009.04335} {arXiv:2009.04335 [gr-qc]} \BibitemShut
  {NoStop}%
\bibitem [{\citenamefont {Milgrom}(1983{\natexlab{a}})}]{Milgrom:1983ca}%
  \BibitemOpen
  \bibfield  {author} {\bibinfo {author} {\bibfnamefont {M.}~\bibnamefont
  {Milgrom}},\ }\href {\doibase 10.1086/161130} {\bibfield  {journal} {\bibinfo
   {journal} {Astrophys. J.}\ }\textbf {\bibinfo {volume} {270}},\ \bibinfo
  {pages} {365} (\bibinfo {year} {1983}{\natexlab{a}})}\BibitemShut {NoStop}%
%%CITATION = ASJOA,270,365;%%
\bibitem [{\citenamefont {Milgrom}(1983{\natexlab{b}})}]{Milgrom:1983pn}%
  \BibitemOpen
  \bibfield  {author} {\bibinfo {author} {\bibfnamefont {M.}~\bibnamefont
  {Milgrom}},\ }\href {\doibase 10.1086/161131} {\bibfield  {journal} {\bibinfo
   {journal} {Astrophys. J.}\ }\textbf {\bibinfo {volume} {270}},\ \bibinfo
  {pages} {371} (\bibinfo {year} {1983}{\natexlab{b}})}\BibitemShut {NoStop}%
%%CITATION = ASJOA,270,371;%%
\bibitem [{\citenamefont {Milgrom}(1983{\natexlab{c}})}]{Milgrom:1983zz}%
  \BibitemOpen
  \bibfield  {author} {\bibinfo {author} {\bibfnamefont {M.}~\bibnamefont
  {Milgrom}},\ }\href {\doibase 10.1086/161132} {\bibfield  {journal} {\bibinfo
   {journal} {Astrophys. J.}\ }\textbf {\bibinfo {volume} {270}},\ \bibinfo
  {pages} {384} (\bibinfo {year} {1983}{\natexlab{c}})}\BibitemShut {NoStop}%
%%CITATION = ASJOA,270,384;%%
\bibitem [{\citenamefont {{Sanders}}\ and\ \citenamefont
  {{McGaugh}}(2002)}]{2002ARA&A..40..263S}%
  \BibitemOpen
  \bibfield  {author} {\bibinfo {author} {\bibfnamefont {R.~H.}\ \bibnamefont
  {{Sanders}}}\ and\ \bibinfo {author} {\bibfnamefont {S.~S.}\ \bibnamefont
  {{McGaugh}}},\ }\href {\doibase 10.1146/annurev.astro.40.060401.093923}
  {\bibfield  {journal} {\bibinfo  {journal} {Annu. Rev. Astron. Astr.}\
  }\textbf {\bibinfo {volume} {40}},\ \bibinfo {pages} {263} (\bibinfo {year}
  {2002})},\ \Eprint {http://arxiv.org/abs/astro-ph/0204521}
  {arXiv:astro-ph/0204521 [astro-ph]} \BibitemShut {NoStop}%
\bibitem [{\citenamefont {Famaey}\ and\ \citenamefont
  {McGaugh}(2012)}]{Famaey:2011kh}%
  \BibitemOpen
  \bibfield  {author} {\bibinfo {author} {\bibfnamefont {B.}~\bibnamefont
  {Famaey}}\ and\ \bibinfo {author} {\bibfnamefont {S.}~\bibnamefont
  {McGaugh}},\ }\href {\doibase 10.12942/lrr-2012-10} {\bibfield  {journal}
  {\bibinfo  {journal} {Living Rev. Rel.}\ }\textbf {\bibinfo {volume} {15}},\
  \bibinfo {pages} {10} (\bibinfo {year} {2012})},\ \Eprint
  {http://arxiv.org/abs/1112.3960} {arXiv:1112.3960 [astro-ph.CO]} \BibitemShut
  {NoStop}%
%%CITATION = ARXIV:1112.3960;%%
\bibitem [{\citenamefont {{Milgrom}}(2014)}]{2014SchpJ...931410M}%
  \BibitemOpen
  \bibfield  {author} {\bibinfo {author} {\bibfnamefont {M.}~\bibnamefont
  {{Milgrom}}},\ }\href {\doibase 10.4249/scholarpedia.31410} {\bibfield
  {journal} {\bibinfo  {journal} {Scholarpedia}\ }\textbf {\bibinfo {volume}
  {9}},\ \bibinfo {pages} {31410} (\bibinfo {year} {2014})}\BibitemShut
  {NoStop}%
\bibitem [{\citenamefont {Merritt}(2020)}]{Merritt:2020pwe}%
  \BibitemOpen
  \bibfield  {author} {\bibinfo {author} {\bibfnamefont {D.}~\bibnamefont
  {Merritt}},\ }\href {\doibase 10.1017/9781108610926} {\emph {\bibinfo {title}
  {{A Philosophical Approach to MOND}}}}\ (\bibinfo  {publisher} {Cambridge
  University Press},\ \bibinfo {year} {2020})\BibitemShut {NoStop}%
\bibitem [{\citenamefont {McGaugh}\ \emph {et~al.}(2016)\citenamefont
  {McGaugh}, \citenamefont {Lelli},\ and\ \citenamefont
  {Schombert}}]{McGaugh:2016leg}%
  \BibitemOpen
  \bibfield  {author} {\bibinfo {author} {\bibfnamefont {S.}~\bibnamefont
  {McGaugh}}, \bibinfo {author} {\bibfnamefont {F.}~\bibnamefont {Lelli}}, \
  and\ \bibinfo {author} {\bibfnamefont {J.}~\bibnamefont {Schombert}},\ }\href
  {\doibase 10.1103/PhysRevLett.117.201101} {\bibfield  {journal} {\bibinfo
  {journal} {Phys. Rev. Lett.}\ }\textbf {\bibinfo {volume} {117}},\ \bibinfo
  {pages} {201101} (\bibinfo {year} {2016})},\ \Eprint
  {http://arxiv.org/abs/1609.05917} {arXiv:1609.05917 [astro-ph.GA]}
  \BibitemShut {NoStop}%
%%CITATION = ARXIV:1609.05917;%%
\bibitem [{\citenamefont {Lelli}\ \emph {et~al.}(2017)\citenamefont {Lelli},
  \citenamefont {McGaugh}, \citenamefont {Schombert},\ and\ \citenamefont
  {Pawlowski}}]{Lelli:2017vgz}%
  \BibitemOpen
  \bibfield  {author} {\bibinfo {author} {\bibfnamefont {F.}~\bibnamefont
  {Lelli}}, \bibinfo {author} {\bibfnamefont {S.~S.}\ \bibnamefont {McGaugh}},
  \bibinfo {author} {\bibfnamefont {J.~M.}\ \bibnamefont {Schombert}}, \ and\
  \bibinfo {author} {\bibfnamefont {M.~S.}\ \bibnamefont {Pawlowski}},\ }\href
  {\doibase 10.3847/1538-4357/836/2/152} {\bibfield  {journal} {\bibinfo
  {journal} {Astrophys. J.}\ }\textbf {\bibinfo {volume} {836}},\ \bibinfo
  {pages} {152} (\bibinfo {year} {2017})},\ \Eprint
  {http://arxiv.org/abs/1610.08981} {arXiv:1610.08981 [astro-ph.GA]}
  \BibitemShut {NoStop}%
%%CITATION = ARXIV:1610.08981;%%
\bibitem [{\citenamefont {Chae}\ \emph {et~al.}(2020)\citenamefont {Chae},
  \citenamefont {Lelli}, \citenamefont {Desmond}, \citenamefont {McGaugh},
  \citenamefont {Li},\ and\ \citenamefont {Schombert}}]{Chae:2020omu}%
  \BibitemOpen
  \bibfield  {author} {\bibinfo {author} {\bibfnamefont {K.-H.}\ \bibnamefont
  {Chae}}, \bibinfo {author} {\bibfnamefont {F.}~\bibnamefont {Lelli}},
  \bibinfo {author} {\bibfnamefont {H.}~\bibnamefont {Desmond}}, \bibinfo
  {author} {\bibfnamefont {S.~S.}\ \bibnamefont {McGaugh}}, \bibinfo {author}
  {\bibfnamefont {P.}~\bibnamefont {Li}}, \ and\ \bibinfo {author}
  {\bibfnamefont {J.~M.}\ \bibnamefont {Schombert}},\ }\href {\doibase
  10.3847/1538-4357/abbb96} {\bibfield  {journal} {\bibinfo  {journal}
  {Astrophys. J.}\ }\textbf {\bibinfo {volume} {904}},\ \bibinfo {pages} {51}
  (\bibinfo {year} {2020})},\ \Eprint {http://arxiv.org/abs/2009.11525}
  {arXiv:2009.11525 [astro-ph.GA]} \BibitemShut {NoStop}%
\bibitem [{\citenamefont {Bollini}\ and\ \citenamefont
  {Giambiagi}(1972)}]{Bollini:1972ui}%
  \BibitemOpen
  \bibfield  {author} {\bibinfo {author} {\bibfnamefont {C.~G.}\ \bibnamefont
  {Bollini}}\ and\ \bibinfo {author} {\bibfnamefont {J.~J.}\ \bibnamefont
  {Giambiagi}},\ }\href {\doibase 10.1007/BF02895558} {\bibfield  {journal}
  {\bibinfo  {journal} {Nuovo Cim.}\ }\textbf {\bibinfo {volume} {B12}},\
  \bibinfo {pages} {20} (\bibinfo {year} {1972})}\BibitemShut {NoStop}%
%%CITATION = NUCIA,B12,20;%%
\bibitem [{\citenamefont {'t~Hooft}\ and\ \citenamefont
  {Veltman}(1972)}]{tHooft:1972tcz}%
  \BibitemOpen
  \bibfield  {author} {\bibinfo {author} {\bibfnamefont {G.}~\bibnamefont
  {'t~Hooft}}\ and\ \bibinfo {author} {\bibfnamefont {M.~J.~G.}\ \bibnamefont
  {Veltman}},\ }\href {\doibase 10.1016/0550-3213(72)90279-9} {\bibfield
  {journal} {\bibinfo  {journal} {Nucl. Phys.}\ }\textbf {\bibinfo {volume}
  {B44}},\ \bibinfo {pages} {189} (\bibinfo {year} {1972})}\BibitemShut
  {NoStop}%
%%CITATION = NUPHA,B44,189;%%
\bibitem [{\citenamefont {Wilson}(1973)}]{Wilson:1972cf}%
  \BibitemOpen
  \bibfield  {author} {\bibinfo {author} {\bibfnamefont {K.~G.}\ \bibnamefont
  {Wilson}},\ }\href {\doibase 10.1103/PhysRevD.7.2911} {\bibfield  {journal}
  {\bibinfo  {journal} {Phys. Rev.}\ }\textbf {\bibinfo {volume} {D7}},\
  \bibinfo {pages} {2911} (\bibinfo {year} {1973})}\BibitemShut {NoStop}%
%%CITATION = PHRVA,D7,2911;%%
\bibitem [{\citenamefont {{Svozil}}(1987)}]{1987JPhA...20.3861S}%
  \BibitemOpen
  \bibfield  {author} {\bibinfo {author} {\bibfnamefont {K.}~\bibnamefont
  {{Svozil}}},\ }\href {\doibase 10.1088/0305-4470/20/12/033} {\bibfield
  {journal} {\bibinfo  {journal} {Journal of Physics A Mathematical General}\
  }\textbf {\bibinfo {volume} {20}},\ \bibinfo {pages} {3861} (\bibinfo {year}
  {1987})}\BibitemShut {NoStop}%
\bibitem [{\citenamefont {Stillinger}(1977)}]{doi:10.1063/1.523395}%
  \BibitemOpen
  \bibfield  {author} {\bibinfo {author} {\bibfnamefont {F.~H.}\ \bibnamefont
  {Stillinger}},\ }\href {\doibase 10.1063/1.523395} {\bibfield  {journal}
  {\bibinfo  {journal} {Journal of Mathematical Physics}\ }\textbf {\bibinfo
  {volume} {18}},\ \bibinfo {pages} {1224} (\bibinfo {year} {1977})},\ \Eprint
  {http://arxiv.org/abs/https://doi.org/10.1063/1.523395}
  {https://doi.org/10.1063/1.523395} \BibitemShut {NoStop}%
\bibitem [{\citenamefont {Palmer}\ and\ \citenamefont
  {Stavrinou}(2004)}]{Palmer_2004}%
  \BibitemOpen
  \bibfield  {author} {\bibinfo {author} {\bibfnamefont {C.}~\bibnamefont
  {Palmer}}\ and\ \bibinfo {author} {\bibfnamefont {P.~N.}\ \bibnamefont
  {Stavrinou}},\ }\href {\doibase 10.1088/0305-4470/37/27/009} {\bibfield
  {journal} {\bibinfo  {journal} {Journal of Physics A: Mathematical and
  General}\ }\textbf {\bibinfo {volume} {37}},\ \bibinfo {pages} {6987}
  (\bibinfo {year} {2004})}\BibitemShut {NoStop}%
\bibitem [{\citenamefont {Bekenstein}\ and\ \citenamefont
  {Milgrom}(1984)}]{Bekenstein:1984tv}%
  \BibitemOpen
  \bibfield  {author} {\bibinfo {author} {\bibfnamefont {J.}~\bibnamefont
  {Bekenstein}}\ and\ \bibinfo {author} {\bibfnamefont {M.}~\bibnamefont
  {Milgrom}},\ }\href {\doibase 10.1086/162570} {\bibfield  {journal} {\bibinfo
   {journal} {Astrophys. J.}\ }\textbf {\bibinfo {volume} {286}},\ \bibinfo
  {pages} {7} (\bibinfo {year} {1984})}\BibitemShut {NoStop}%
%%CITATION = ASJOA,286,7;%%
\bibitem [{\citenamefont {Petersen}\ and\ \citenamefont
  {Lelli}(2020)}]{Petersen:2020vks}%
  \BibitemOpen
  \bibfield  {author} {\bibinfo {author} {\bibfnamefont {J.}~\bibnamefont
  {Petersen}}\ and\ \bibinfo {author} {\bibfnamefont {F.}~\bibnamefont
  {Lelli}},\ }\href {\doibase 10.1051/0004-6361/201936964} {\bibfield
  {journal} {\bibinfo  {journal} {Astron. Astrophys.}\ }\textbf {\bibinfo
  {volume} {636}},\ \bibinfo {pages} {A56} (\bibinfo {year} {2020})},\ \Eprint
  {http://arxiv.org/abs/2001.03348} {arXiv:2001.03348 [astro-ph.GA]}
  \BibitemShut {NoStop}%
\bibitem [{\citenamefont {Milgrom}(2012)}]{Milgrom:2012rk}%
  \BibitemOpen
  \bibfield  {author} {\bibinfo {author} {\bibfnamefont {M.}~\bibnamefont
  {Milgrom}},\ }\href {\doibase 10.1103/PhysRevLett.109.251103} {\bibfield
  {journal} {\bibinfo  {journal} {Phys.\ Rev.\ Lett.}\ }\textbf {\bibinfo
  {volume} {109}},\ \bibinfo {pages} {251103} (\bibinfo {year} {2012})},\
  \Eprint {http://arxiv.org/abs/1211.4899} {arXiv:1211.4899 [astro-ph.CO]}
  \BibitemShut {NoStop}%
\bibitem [{\citenamefont {McGaugh}(2008)}]{McGaugh:2008nc}%
  \BibitemOpen
  \bibfield  {author} {\bibinfo {author} {\bibfnamefont {S.}~\bibnamefont
  {McGaugh}},\ }\href {\doibase 10.1086/589148} {\bibfield  {journal} {\bibinfo
   {journal} {Astrophys. J.}\ }\textbf {\bibinfo {volume} {683}},\ \bibinfo
  {pages} {137} (\bibinfo {year} {2008})},\ \Eprint
  {http://arxiv.org/abs/0804.1314} {arXiv:0804.1314 [astro-ph]} \BibitemShut
  {NoStop}%
%%CITATION = ARXIV:0804.1314;%%
\bibitem [{\citenamefont {Lelli}\ \emph {et~al.}(2016)\citenamefont {Lelli},
  \citenamefont {McGaugh},\ and\ \citenamefont {Schombert}}]{Lelli:2016zqa}%
  \BibitemOpen
  \bibfield  {author} {\bibinfo {author} {\bibfnamefont {F.}~\bibnamefont
  {Lelli}}, \bibinfo {author} {\bibfnamefont {S.~S.}\ \bibnamefont {McGaugh}},
  \ and\ \bibinfo {author} {\bibfnamefont {J.~M.}\ \bibnamefont {Schombert}},\
  }\href {\doibase 10.3847/0004-6256/152/6/157} {\bibfield  {journal} {\bibinfo
   {journal} {Astron. J.}\ }\textbf {\bibinfo {volume} {152}},\ \bibinfo
  {pages} {157} (\bibinfo {year} {2016})},\ \Eprint
  {http://arxiv.org/abs/1606.09251} {arXiv:1606.09251 [astro-ph.GA]}
  \BibitemShut {NoStop}%
%%CITATION = ARXIV:1606.09251;%%
\bibitem [{\citenamefont {Li}\ \emph {et~al.}(2018)\citenamefont {Li},
  \citenamefont {Lelli}, \citenamefont {McGaugh},\ and\ \citenamefont
  {Schombert}}]{Li:2018tdo}%
  \BibitemOpen
  \bibfield  {author} {\bibinfo {author} {\bibfnamefont {P.}~\bibnamefont
  {Li}}, \bibinfo {author} {\bibfnamefont {F.}~\bibnamefont {Lelli}}, \bibinfo
  {author} {\bibfnamefont {S.}~\bibnamefont {McGaugh}}, \ and\ \bibinfo
  {author} {\bibfnamefont {J.}~\bibnamefont {Schombert}},\ }\href {\doibase
  10.1051/0004-6361/201732547} {\bibfield  {journal} {\bibinfo  {journal}
  {Astron. Astrophys.}\ }\textbf {\bibinfo {volume} {615}},\ \bibinfo {pages}
  {A3} (\bibinfo {year} {2018})},\ \Eprint {http://arxiv.org/abs/1803.00022}
  {arXiv:1803.00022 [astro-ph.GA]} \BibitemShut {NoStop}%
%%CITATION = ARXIV:1803.00022;%%
\bibitem [{\citenamefont {{Binney}}\ and\ \citenamefont
  {{Tremaine}}(2008)}]{2008gady.book.....B}%
  \BibitemOpen
  \bibfield  {author} {\bibinfo {author} {\bibfnamefont {J.}~\bibnamefont
  {{Binney}}}\ and\ \bibinfo {author} {\bibfnamefont {S.}~\bibnamefont
  {{Tremaine}}},\ }\href@noop {} {\emph {\bibinfo {title} {Galactic Dynamics:
  Second Edition, by James Binney and Scott Tremaine. ISBN 978-0-691-13026-2
  (HB). Published by Princeton University Press, Princeton, NJ USA, 2008.}}}\
  (\bibinfo {year} {2008})\BibitemShut {NoStop}%
\bibitem [{\citenamefont {Mannheim}(2006)}]{Mannheim:2005bfa}%
  \BibitemOpen
  \bibfield  {author} {\bibinfo {author} {\bibfnamefont {P.~D.}\ \bibnamefont
  {Mannheim}},\ }\href {\doibase 10.1016/j.ppnp.2005.08.001} {\bibfield
  {journal} {\bibinfo  {journal} {Prog. Part. Nucl. Phys.}\ }\textbf {\bibinfo
  {volume} {56}},\ \bibinfo {pages} {340} (\bibinfo {year} {2006})},\ \Eprint
  {http://arxiv.org/abs/astro-ph/0505266} {arXiv:astro-ph/0505266 [astro-ph]}
  \BibitemShut {NoStop}%
%%CITATION = ASTRO-PH/0505266;%%
\bibitem [{\citenamefont {Jackson}(1998)}]{Jackson:1998nia}%
  \BibitemOpen
  \bibfield  {author} {\bibinfo {author} {\bibfnamefont {J.~D.}\ \bibnamefont
  {Jackson}},\ }\href@noop {} {\emph {\bibinfo {title} {{Classical
  Electrodynamics}}}}\ (\bibinfo  {publisher} {Wiley},\ \bibinfo {year}
  {1998})\BibitemShut {NoStop}%
\bibitem [{\citenamefont {{Casertano}}(1983)}]{1983MNRAS.203..735C}%
  \BibitemOpen
  \bibfield  {author} {\bibinfo {author} {\bibfnamefont {S.}~\bibnamefont
  {{Casertano}}},\ }\href {\doibase 10.1093/mnras/203.3.735} {\bibfield
  {journal} {\bibinfo  {journal} {Mon. Not. Roy. Astron. Soc.}\ }\textbf
  {\bibinfo {volume} {203}},\ \bibinfo {pages} {735} (\bibinfo {year}
  {1983})}\BibitemShut {NoStop}%
\bibitem [{\citenamefont {{Cohl}}\ and\ \citenamefont
  {{Kalnins}}(2012)}]{2012JPhA...45n5206C}%
  \BibitemOpen
  \bibfield  {author} {\bibinfo {author} {\bibfnamefont {H.~S.}\ \bibnamefont
  {{Cohl}}}\ and\ \bibinfo {author} {\bibfnamefont {E.~G.}\ \bibnamefont
  {{Kalnins}}},\ }\href {\doibase 10.1088/1751-8113/45/14/145206} {\bibfield
  {journal} {\bibinfo  {journal} {Journal of Physics A Mathematical General}\
  }\textbf {\bibinfo {volume} {45}},\ \bibinfo {eid} {145206} (\bibinfo {year}
  {2012})},\ \Eprint {http://arxiv.org/abs/1105.0386} {arXiv:1105.0386
  [math-ph]} \BibitemShut {NoStop}%
\bibitem [{{\relax DLMF}()}]{NIST:DLMF}%
  \BibitemOpen
  {\relax DLMF},\ \href {http://dlmf.nist.gov/} {\enquote {\bibinfo {title}
  {{\it NIST Digital Library of Mathematical Functions}},}\ }\bibinfo
  {howpublished} {http://dlmf.nist.gov/, Release 1.0.25 of 2019-12-15},\
  \bibinfo {note} {f.~W.~J. Olver, A.~B. {Olde Daalhuis}, D.~W. Lozier, B.~I.
  Schneider, R.~F. Boisvert, C.~W. Clark, B.~R. Miller, B.~V. Saunders, H.~S.
  Cohl, and M.~A. McClain, eds.}\BibitemShut {Stop}%
\bibitem [{\citenamefont {Cohl}(2013)}]{doi:10.1080/10652469.2012.761613}%
  \BibitemOpen
  \bibfield  {author} {\bibinfo {author} {\bibfnamefont {H.~S.}\ \bibnamefont
  {Cohl}},\ }\href {\doibase 10.1080/10652469.2012.761613} {\bibfield
  {journal} {\bibinfo  {journal} {Integral Transforms and Special Functions}\
  }\textbf {\bibinfo {volume} {24}},\ \bibinfo {pages} {807} (\bibinfo {year}
  {2013})},\ \Eprint
  {http://arxiv.org/abs/https://doi.org/10.1080/10652469.2012.761613}
  {https://doi.org/10.1080/10652469.2012.761613} \BibitemShut {NoStop}%
\bibitem [{\citenamefont {{Cohl}}(2013)}]{2013SIGMA...9..042C}%
  \BibitemOpen
  \bibfield  {author} {\bibinfo {author} {\bibfnamefont {H.~S.}\ \bibnamefont
  {{Cohl}}},\ }\href {\doibase 10.3842/SIGMA.2013.042} {\bibfield  {journal}
  {\bibinfo  {journal} {SIGMA}\ }\textbf {\bibinfo {volume} {9}},\ \bibinfo
  {eid} {042} (\bibinfo {year} {2013})},\ \Eprint
  {http://arxiv.org/abs/1209.6047} {arXiv:1209.6047 [math-ph]} \BibitemShut
  {NoStop}%
\bibitem [{\citenamefont {{Cohl}}\ and\ \citenamefont
  {{Palmer}}(2015)}]{2015SIGMA..11..015C}%
  \BibitemOpen
  \bibfield  {author} {\bibinfo {author} {\bibfnamefont {H.~S.}\ \bibnamefont
  {{Cohl}}}\ and\ \bibinfo {author} {\bibfnamefont {R.~M.}\ \bibnamefont
  {{Palmer}}},\ }\href {\doibase 10.3842/SIGMA.2015.015} {\bibfield  {journal}
  {\bibinfo  {journal} {SIGMA}\ }\textbf {\bibinfo {volume} {11}},\ \bibinfo
  {eid} {015} (\bibinfo {year} {2015})},\ \Eprint
  {http://arxiv.org/abs/1405.4847} {arXiv:1405.4847 [math.CA]} \BibitemShut
  {NoStop}%
\bibitem [{\citenamefont {Tarasov}(2014)}]{Tarasov:2014fda}%
  \BibitemOpen
  \bibfield  {author} {\bibinfo {author} {\bibfnamefont {V.~E.}\ \bibnamefont
  {Tarasov}},\ }\href {\doibase 10.1063/1.4892155} {\bibfield  {journal}
  {\bibinfo  {journal} {J. Math. Phys.}\ }\textbf {\bibinfo {volume} {55}},\
  \bibinfo {pages} {083510} (\bibinfo {year} {2014})},\ \Eprint
  {http://arxiv.org/abs/1503.02392} {arXiv:1503.02392 [math-ph]} \BibitemShut
  {NoStop}%
%%CITATION = ARXIV:1503.02392;%%
\bibitem [{\citenamefont {Tarasov}(2015)}]{TARASOV2015360}%
  \BibitemOpen
  \bibfield  {author} {\bibinfo {author} {\bibfnamefont {V.~E.}\ \bibnamefont
  {Tarasov}},\ }\href {\doibase https://doi.org/10.1016/j.cnsns.2014.05.025}
  {\bibfield  {journal} {\bibinfo  {journal} {Communications in Nonlinear
  Science and Numerical Simulation}\ }\textbf {\bibinfo {volume} {20}},\
  \bibinfo {pages} {360 } (\bibinfo {year} {2015})}\BibitemShut {NoStop}%
\bibitem [{\citenamefont {{Bershady}}\ \emph {et~al.}(2010)\citenamefont
  {{Bershady}}, \citenamefont {{Verheijen}}, \citenamefont {{Westfall}},
  \citenamefont {{Andersen}}, \citenamefont {{Swaters}},\ and\ \citenamefont
  {{Martinsson}}}]{2010ApJ...716..234B}%
  \BibitemOpen
  \bibfield  {author} {\bibinfo {author} {\bibfnamefont {M.~A.}\ \bibnamefont
  {{Bershady}}}, \bibinfo {author} {\bibfnamefont {M.~A.~W.}\ \bibnamefont
  {{Verheijen}}}, \bibinfo {author} {\bibfnamefont {K.~B.}\ \bibnamefont
  {{Westfall}}}, \bibinfo {author} {\bibfnamefont {D.~R.}\ \bibnamefont
  {{Andersen}}}, \bibinfo {author} {\bibfnamefont {R.~A.}\ \bibnamefont
  {{Swaters}}}, \ and\ \bibinfo {author} {\bibfnamefont {T.}~\bibnamefont
  {{Martinsson}}},\ }\href {\doibase 10.1088/0004-637X/716/1/234} {\bibfield
  {journal} {\bibinfo  {journal} {\apj}\ }\textbf {\bibinfo {volume} {716}},\
  \bibinfo {pages} {234} (\bibinfo {year} {2010})},\ \Eprint
  {http://arxiv.org/abs/1004.5043} {arXiv:1004.5043 [astro-ph.CO]} \BibitemShut
  {NoStop}%
\bibitem [{\citenamefont {Carlip}(2019)}]{Carlip:2019onx}%
  \BibitemOpen
  \bibfield  {author} {\bibinfo {author} {\bibfnamefont {S.}~\bibnamefont
  {Carlip}},\ }\href {\doibase 10.3390/universe5030083} {\bibfield  {journal}
  {\bibinfo  {journal} {Universe}\ }\textbf {\bibinfo {volume} {5}},\ \bibinfo
  {pages} {83} (\bibinfo {year} {2019})},\ \Eprint
  {http://arxiv.org/abs/1904.04379} {arXiv:1904.04379 [gr-qc]} \BibitemShut
  {NoStop}%
\bibitem [{\citenamefont {Varieschi}(2020{\natexlab{c}})}]{Varieschi:2020hvp}%
  \BibitemOpen
  \bibfield  {author} {\bibinfo {author} {\bibfnamefont {G.~U.}\ \bibnamefont
  {Varieschi}},\ }\href@noop {} {\  (\bibinfo {year} {2020}{\natexlab{c}})},\
  \Eprint {http://arxiv.org/abs/2011.04911} {arXiv:2011.04911 [gr-qc]}
  \BibitemShut {NoStop}%
\bibitem [{\citenamefont {Lelli}(2020)}]{Lelli:2020pri}%
  \BibitemOpen
  \bibfield  {author} {\bibinfo {author} {\bibfnamefont {F.}~\bibnamefont
  {Lelli}},\ }\href@noop {} {\bibfield  {journal} {\bibinfo  {journal} {private
  communication}\ } (\bibinfo {year} {2020})}\BibitemShut {NoStop}%
\bibitem [{\citenamefont {Will}(2014)}]{Will:2014kxa}%
  \BibitemOpen
  \bibfield  {author} {\bibinfo {author} {\bibfnamefont {C.~M.}\ \bibnamefont
  {Will}},\ }\href {\doibase 10.12942/lrr-2014-4} {\bibfield  {journal}
  {\bibinfo  {journal} {Living Rev. Rel.}\ }\textbf {\bibinfo {volume} {17}},\
  \bibinfo {pages} {4} (\bibinfo {year} {2014})},\ \Eprint
  {http://arxiv.org/abs/1403.7377} {arXiv:1403.7377 [gr-qc]} \BibitemShut
  {NoStop}%
\end{thebibliography}%

\end{document}